\documentclass[aps,prd,twocolumn,nofootinbib,longbibliography]{revtex4-2}
\pdfoutput=1
\usepackage{amsmath,amssymb}
\usepackage{graphicx}
\graphicspath{{./}}
\usepackage{bm}
\usepackage[hidelinks]{hyperref}
\usepackage{balance}

\newcommand{\Pz}{P_\zeta}
\newcommand{\PBD}{P_\zeta^{\rm BD}}
\newcommand{\kstar}{k_\ast}
\newcommand{\Dk}{\Delta k}
\newcommand{\Mpl}{M_{\rm Pl}}
\newcommand{\SN}{\mathcal{S}_N}
\newcommand{\de}{\mathrm{d}}
\newcommand{\order}{\mathcal{O}}

\begin{document}

\title{Superoscillatory initial states during inflation: theory, CMB constraints, and prospects for galaxy clustering}

\author{Ali Nayeri}
\email{nayeri@mit.edu}
\affiliation{Clear Quantum Corporation, Lewes, Delaware 19958, USA and\\
Ordinal Research Institute, Wilmington, DE 19801, USA}
\date{June 5, 2026}

\begin{abstract}
We construct an explicit boundary-action realization of superoscillatory
initial states (SIS) for inflation, in which quantum interference within
a band-limited initial wavefunctional generates a spectrally localized
Bogoliubov excitation with a rapidly winding phase. Starting from a
quadratic boundary term on the initial time surface, we derive the
Bogoliubov coefficients and the resulting primordial curvature spectrum,
obtaining a localized oscillatory feature fixed by the superoscillatory
parameters $(a,N)$ rather than imposed phenomenologically. We compute the
projection of this feature onto CMB angular power spectra and show that
transfer-function smearing strongly suppresses the oscillatory component;
full CAMB calculations confirm the qualitative effect and show that a
simple Gaussian approximation overestimates the peak signal by about a
factor of three. Using Planck 2018 TT data, we obtain an indicative
matched-filter bound $\lambda \lesssim 0.05$ for a representative feature
centered near the first acoustic peak, $\Delta k/k_* = 0.05$ at
$k_* = 1.45\times 10^{-2}\,\mathrm{Mpc}^{-1}$.
We further derive correlated predictions for polarization and the
bispectrum, identify structural constraints that distinguish SIS from
generic excited-state models, and show that galaxy clustering provides a
qualitatively more powerful probe because it preserves the full
oscillatory structure that CMB projection suppresses. This framework
provides a concrete and testable realization of how initial-state quantum
interference can imprint itself on cosmological observables.
\end{abstract}

\maketitle

\section{Introduction}
\label{sec:intro}

Inflation provides a compelling origin for the nearly scale-invariant
primordial perturbations observed in the cosmic microwave background
(CMB)~\cite{Guth1981,Linde1982,Linde1983,Albrecht1982,Planck2018params}.
The standard assumption is that inflaton fluctuations begin in the
Bunch--Davies (BD) vacuum, yielding a smooth curvature spectrum
$\PBD(k)$.  Departures from BD---through excited or non-adiabatic
initial states---can produce oscillatory features in the primordial
spectrum~\cite{Holman2008,Agullo2011,Chen2010,Danielsson2002,BrandenbergerMartin2001}.
The trans-Planckian problem of inflation~\cite{BrandenbergerMartin2013,Easther2001,Easther2002}
provides additional motivation for studying such departures, since modes
observed in the CMB today may have originated with wavelengths shorter
than the Planck length at the onset of inflation.

We consider a distinct quantum-mechanical mechanism: a superoscillatory
initial state (SIS), in which quantum interference among band-limited
modes produces localized phase gradients exceeding the global
bandwidth~\cite{Aharonov1994,Berry1994,BerryPopescu2006,AharonovReview2017}.
Superoscillations---a phenomenon in which a band-limited function
oscillates locally faster than its fastest Fourier component---have been
realized experimentally in optics~\cite{Huang2007} and have deep
connections to weak measurements in quantum
mechanics~\cite{AharonovVaidman1990,Berry2019}.
The key phenomenological consequence for inflation is spectral
localization: the excitation occupies a narrow band in $k$ and therefore
projects to a narrow band in multipole $\ell$.

The aims of this paper are fourfold.  First, we provide an explicit
boundary-action construction that derives the SIS Bogoliubov coefficients
from a well-defined initial-state kernel, replacing a purely
phenomenological parametrization of $\beta_k$.  The resulting framework
inherits the rich mathematical structure of superoscillation
theory---band-limitedness, the Gaussian dominance condition, binomial
harmonic decomposition---and translates it into a tightly constrained
inflationary phenomenology.  Second, we carry out a
careful projection of the primordial feature onto CMB angular power
spectra, accounting for the transfer-function smearing that is critical
for assessing observability, and calibrate the analytic results against
full Boltzmann computation.  Third, we confront the SIS template with
Planck 2018 data~\cite{Planck2018spectra}, deriving indicative bounds on the
excitation amplitude.  Fourth, we identify galaxy clustering as a
qualitatively superior probe of the SIS feature---free of the
transfer-function smearing that limits the CMB---and present a
preliminary Fisher estimate exploring whether DESI~\cite{DESI2024} could
offer competitive sensitivity.

Within the present phenomenological implementation, the superoscillatory
construction fixes the local shape of the excitation but not its central
comoving scale $\kstar$.  We therefore treat $\kstar$ as a scan parameter
in the observational analysis and quote
$\kstar = 1.45\times10^{-2}$~Mpc$^{-1}$ as a fiducial benchmark because
it projects near the first acoustic peak, where the CMB temperature data
have high sensitivity to localized features.

The paper is organized as follows.
Section~\ref{sec:boundary} derives the SIS state from a boundary action
and obtains the modified primordial spectrum.
Section~\ref{sec:cmb} computes the CMB projection including
transfer-function smearing and polarization predictions.
Section~\ref{sec:constraints} presents observational constraints from
Planck TT data and CMB forecasts.
Section~\ref{sec:beyond} derives predictions for the bispectrum and
tensor sector.
Section~\ref{sec:distinction} identifies the structural constraints
that distinguish SIS from generic excited-state models.
Section~\ref{sec:desi} explores the prospects for detecting the SIS
feature in galaxy clustering, with a preliminary Fisher estimate for
DESI.
Section~\ref{sec:discussion} discusses implications and outlook.
Appendix~\ref{app:wronskian} verifies the Wronskian normalization and
provides intermediate steps for the boundary-condition derivation.
Appendix~\ref{app:convolution} derives the transfer-function convolution
formula in detail.

\section{Boundary-action derivation of the SIS state}
\label{sec:boundary}

\subsection{General Gaussian initial state}
\label{sec:gaussian}

Scalar curvature perturbations are described by the Mukhanov--Sasaki
variable $v = z\zeta$ with action
\begin{equation}
S = \frac{1}{2}\int \de\tau\,\de^3 x
\left[v'^2 - (\nabla v)^2 + \frac{z''}{z}\,v^2\right],
\label{eq:MS_action}
\end{equation}
where $z = a\dot\phi/H$ and primes denote conformal-time derivatives.
In quasi--de~Sitter, $z''/z \simeq 2/\tau^2$, and the Bunch--Davies mode
function is
\begin{equation}
u_k(\tau) = \frac{1}{\sqrt{2k}}
\left(1 - \frac{i}{k\tau}\right)e^{-ik\tau}.
\label{eq:BD_mode}
\end{equation}
A general Gaussian initial state can be written as
\begin{equation}
v_k(\tau) = \alpha_k\,u_k(\tau) + \beta_k\,u_k^*(\tau),
\qquad |\alpha_k|^2 - |\beta_k|^2 = 1,
\label{eq:Bogoliubov}
\end{equation}
giving the primordial curvature spectrum
\begin{equation}
\Pz(k) = \PBD(k)\,|\alpha_k + \beta_k|^2.
\label{eq:Pz_general}
\end{equation}

\subsection{Boundary action and Robin condition}
\label{sec:robin}

We supplement the bulk action~\eqref{eq:MS_action} with a quadratic
boundary term at the initial time $\tau = \tau_*$:
\begin{equation}
S_{\rm bdy} = \frac{1}{2}\int_{\tau=\tau_*}
\frac{\de^3 k}{(2\pi)^3}\,\kappa_k\,v_{\mathbf{k}}\,v_{-\mathbf{k}},
\label{eq:S_boundary}
\end{equation}
where $\kappa_k = \kappa(k)$ is an isotropic kernel.  Varying the total
action $S + S_{\rm bdy}$ and requiring the boundary variation to vanish
gives the Robin boundary condition
\begin{equation}
\bigl[v_k'(\tau) - \kappa_k\,v_k(\tau)\bigr]_{\tau=\tau_*} = 0.
\label{eq:Robin_bc}
\end{equation}
Substituting the mode expansion~\eqref{eq:Bogoliubov} into
eq.~\eqref{eq:Robin_bc} yields
\begin{equation}
\frac{\beta_k}{\alpha_k}
= -\frac{u_k'(\tau_*) - \kappa_k\,u_k(\tau_*)}
{u_k'^*(\tau_*) - \kappa_k\,u_k^*(\tau_*)}.
\label{eq:beta_alpha_exact}
\end{equation}

The Bunch--Davies vacuum corresponds to
$\kappa_k^{\rm BD} = u_k'(\tau_*)/u_k(\tau_*)$, for which
$\beta_k = 0$.  Writing
$\kappa_k = \kappa_k^{\rm BD} + \delta\kappa_k$ and using the
Wronskian normalization $u_k u_k'^* - u_k^* u_k' = i$, the numerator
of eq.~\eqref{eq:beta_alpha_exact} reduces to
$-\delta\kappa_k\,u_k(\tau_*)$ and the denominator to
$i/u_k(\tau_*) - \delta\kappa_k\,u_k^*(\tau_*)$.  Therefore
\begin{equation}
\frac{\beta_k}{\alpha_k}
= \frac{\delta\kappa_k\,u_k(\tau_*)^2}
{i - \delta\kappa_k\,|u_k(\tau_*)|^2}.
\label{eq:beta_from_kernel}
\end{equation}
For perturbative excitations
$|\delta\kappa_k|\,|u_k(\tau_*)|^2 \ll 1$, this simplifies to
\begin{equation}
\beta_k \simeq -i\,\delta\kappa_k\,u_k(\tau_*)^2.
\label{eq:beta_linear}
\end{equation}

\paragraph{Relation to other initial-state prescriptions.}
The boundary-action framework of eq.~\eqref{eq:S_boundary} encompasses
several initial-state constructions in the literature.  Danielsson's
$\alpha$-vacuum prescription~\cite{Danielsson2002} corresponds to choosing
$\kappa_k$ so that $v_k(\tau_*)$ satisfies the WKB normalization condition
on the initial hypersurface, giving a broadband excitation
$\beta_k \sim 1/(k\tau_*)^2$ that affects all modes uniformly.  The SIS
construction differs in two ways: $\delta\kappa_k$ has narrow spectral
support (localized near $\kstar$), and the superoscillatory phase
structure $\SN$ produces a rapidly winding phase with no counterpart in
the $\alpha$-vacuum.  The general initial-state problem reduces to
specifying the kernel $\kappa_k$, with $\alpha$-vacua (smooth and
broadband) and SIS states (localized and rapidly phased) as special cases
within a unified formalism.

\subsection{Superoscillatory boundary kernel}
\label{sec:so_kernel}

We now choose a specific deformation that produces a superoscillatory
Bogoliubov coefficient.  Define the dimensionless variable
$x \equiv (k - \kstar)/\Dk$ and introduce the standard superoscillatory
function~\cite{Aharonov1994,Berry1994}
\begin{equation}
\SN(x) = \left[\cos\!\left(\frac{x}{N}\right)
+ ia\sin\!\left(\frac{x}{N}\right)\right]^N, \qquad a > 1.
\label{eq:SN_def}
\end{equation}
Its binomial expansion
\begin{equation}
\SN(x) = \sum_{m=0}^{N}\binom{N}{m}
\left(\frac{1+a}{2}\right)^m
\left(\frac{1-a}{2}\right)^{N-m}
e^{i(2m-N)x/N}
\label{eq:SN_binomial}
\end{equation}
shows that all Fourier frequencies satisfy
$\omega_m = (2m-N)/N \in [-1,1]$, so $\SN$ is globally band-limited.
Inside the superoscillatory patch $|x| \ll \sqrt{N}$, however,
\begin{equation}
\SN(x) = \exp\!\left[iax + \order\!\left(\frac{x^2}{N}\right)\right],
\label{eq:SN_patch}
\end{equation}
so the local phase gradient $\de\arg\SN/\de x \simeq a > 1$ exceeds
the global bandwidth.

We choose the boundary deformation
\begin{equation}
\delta\kappa_k = \frac{i\lambda}{u_k(\tau_*)^2}\,
\exp\!\left[-\frac{(k-\kstar)^2}{2\Dk^2}\right]
\SN\!\left(\frac{k-\kstar}{\Dk}\right),
\qquad \lambda \ll 1.
\label{eq:delta_kappa}
\end{equation}
The factor $1/u_k(\tau_*)^2$ is constructed so that the mode-function
dependence cancels in eq.~\eqref{eq:beta_linear}, giving
\begin{equation}
\beta_k \simeq \lambda\,
\exp\!\left[-\frac{(k-\kstar)^2}{2\Dk^2}\right]
\SN\!\left(\frac{k-\kstar}{\Dk}\right).
\label{eq:beta_derived}
\end{equation}
This is an inverse-problem construction within the boundary effective
field theory framework~\cite{Schalm2004,Greene2005,CollinsHolman2005}:
the kernel is chosen to produce the desired $\beta_k$, analogous to
specifying a Wilsonian boundary operator and fixing its coefficient by
matching to a target excitation spectrum.  In the present work we adopt
this as a phenomenological strategy rather than deriving the kernel from
a specific UV completion.  Any boundary kernel that yields a spectrally
localized, rapidly phased Bogoliubov coefficient will produce the same
phenomenology; the specific form of eq.~\eqref{eq:delta_kappa} is a
minimal realization.

\paragraph{Physical interpretation.}
For sub-horizon modes ($|k\tau_*| \gg 1$), $u_k(\tau_*)^2 \approx
e^{-2ik\tau_*}/(2k)$, so $\delta\kappa_k \approx
2ik\lambda\,e^{2ik\tau_*}\,G(k)\,\SN(x)$ where $G(k)$ is the Gaussian
envelope.  The factor $e^{2ik\tau_*}$ is the standard matching phase that
any boundary deformation requires at $\tau_*$; the physical content
resides in the Gaussian $\times\;\SN$ factor.  A natural setting for
such a kernel is a new-physics hypersurface at $\tau_*$ where the
effective field theory of inflation transitions from a UV-complete regime
to the standard slow-roll description.  If this transition is spatially
inhomogeneous---with a localized patch of comoving size
$L_{\rm so} \sim 1/\Dk$ where the transition is anomalously rapid---the
boundary data would carry a superoscillatory imprint: the Gaussian
envelope from the finite extent of the patch, and the rapidly winding
phase from the short-distance structure within it.  At the same time, we
do not claim here to derive such a patch from a concrete brane/boundary
construction, string compactification, or lattice initial state.  The
point of the SIS kernel is instead to provide a controlled EFT
parameterization of what such UV data would have to look like if it were
to generate a superoscillatory excitation.  An explicit microscopic
derivation remains an open problem.

As a proof of principle, one may view the SIS kernel as a localized
version of the non-adiabatic matching familiar from trans-Planckian
initial-state models~\cite{Easther2001,Easther2002}.  In such
constructions, modified short-distance dynamics near a new-physics
hypersurface generates a Bogoliubov coefficient with the same universal
matching phase $e^{2ik\tau_*}$ and an amplitude controlled by the
departure from adiabatic evolution.  If that departure is itself
confined to a finite comoving region or to a narrow band of physical
momenta, the resulting boundary kernel is naturally localized in $k$ and
is well approximated by a smooth envelope multiplying the matching phase.
The SIS ansatz can then be interpreted as the additional statement that
the microscopic boundary data within that localized region are
band-limited, so that their interference produces the superoscillatory
factor $\SN(x)$.  This does not amount to a UV completion, but it gives
a concrete example of how known trans-Planckian EFT logic can generate a
kernel of the general form used here.

\begin{figure*}[t]
\centering
\includegraphics[width=\textwidth]{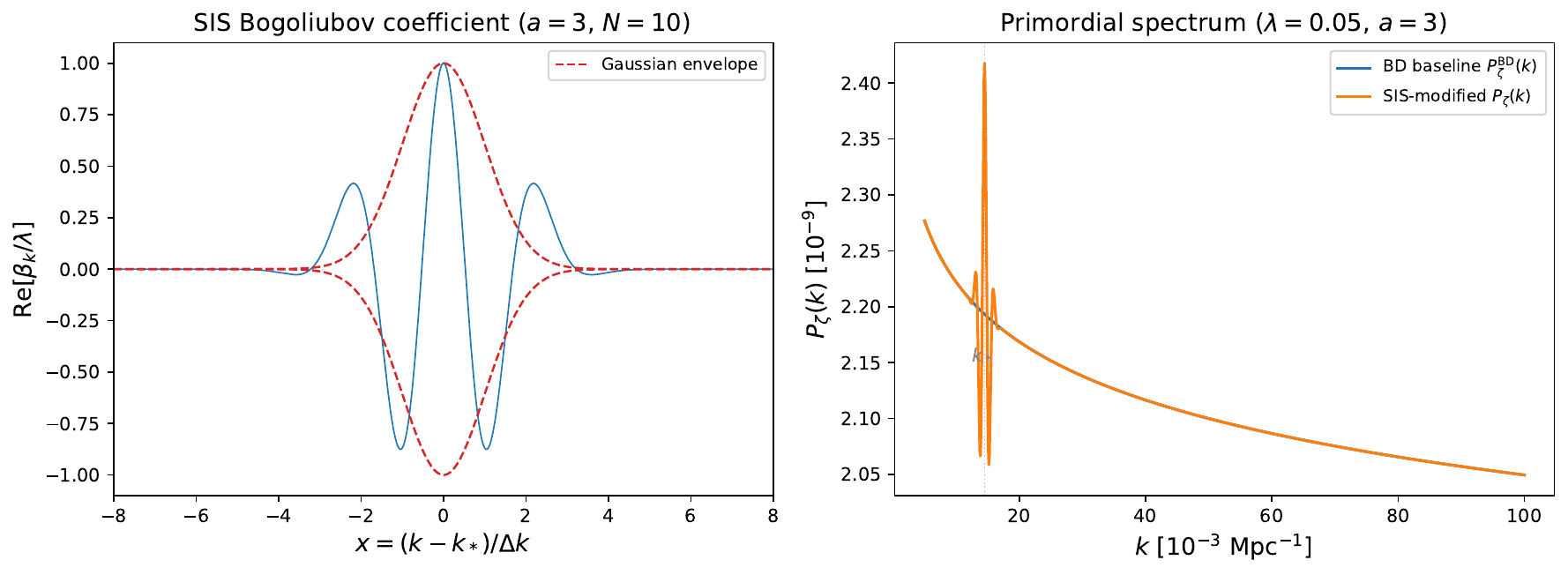}
\caption{SIS mechanism.  Left: real part of the derived Bogoliubov
coefficient $\beta_k/\lambda$ showing the rapidly oscillating
superoscillatory patch inside the Gaussian envelope.
Right: the primordial curvature spectrum with a representative localized
SIS feature
at $\kstar = 1.45\times 10^{-2}$~Mpc$^{-1}$.}
\label{fig:mechanism}
\end{figure*}

Figure~\ref{fig:mechanism} illustrates the
resulting Bogoliubov coefficient and its imprint on the primordial
spectrum.

Inside the superoscillatory patch, eq.~\eqref{eq:beta_derived} reduces to
\begin{equation}
\beta_k \simeq \lambda\,
\exp\!\left[-\frac{(k-\kstar)^2}{2\Dk^2}\right]
\exp\!\left[ia\,\frac{k-\kstar}{\Dk}\right],
\label{eq:beta_patch}
\end{equation}
exhibiting both the Gaussian localization and the rapidly winding phase
that characterize the SIS mechanism.

\subsection{Derived power spectrum}
\label{sec:derived_Pz}

\begin{figure*}[t]
\centering
\includegraphics[width=\textwidth]{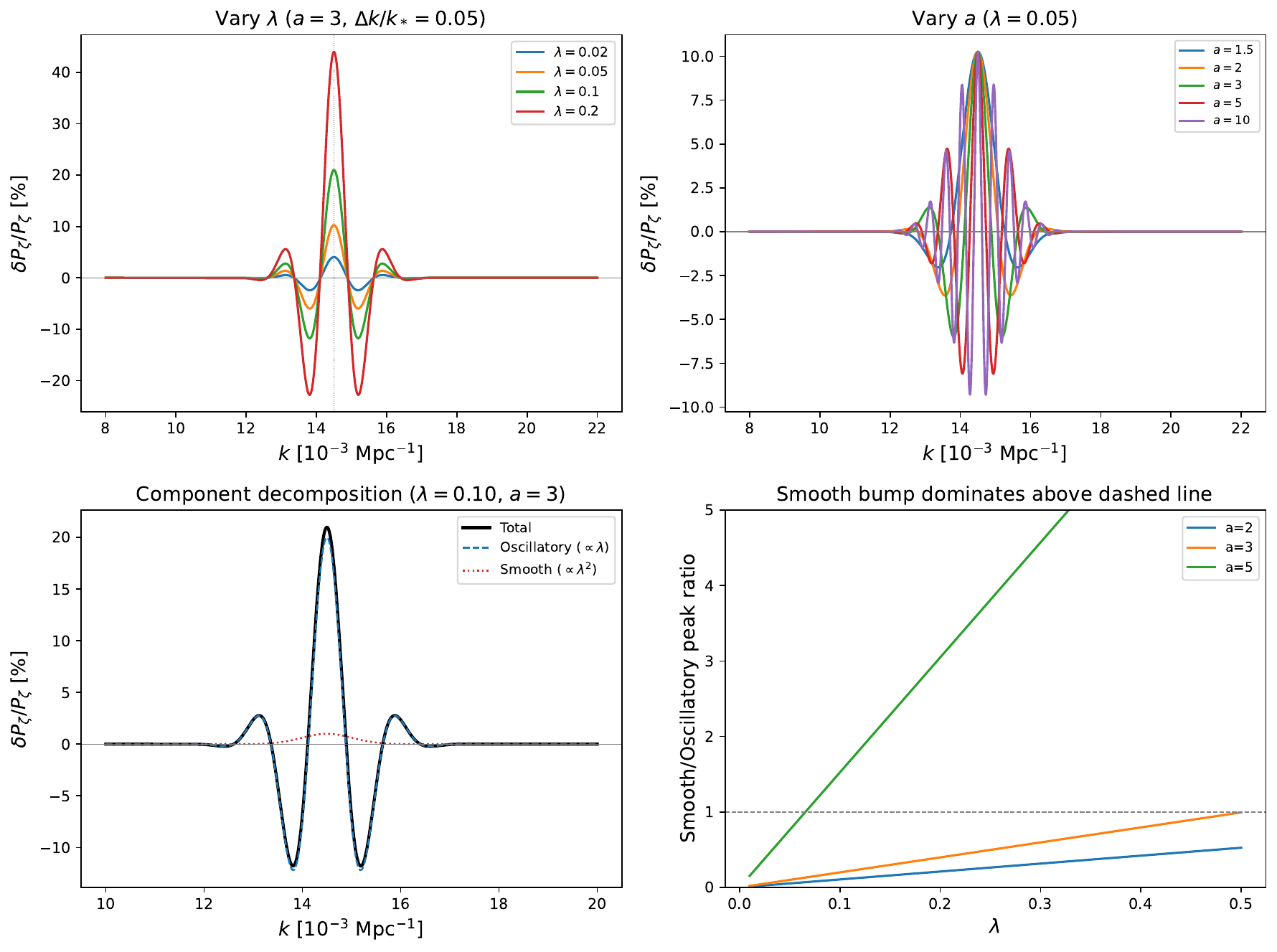}
\caption{Derived SIS power-spectrum template.  Top left: dependence on
$\lambda$ at fixed $a=3$.  Top right: dependence on the superoscillatory
parameter $a$ at fixed $\lambda=0.05$.  Bottom left: decomposition into
oscillatory (linear in $\lambda$) and smooth (quadratic in $\lambda$)
components.  Bottom right: the smooth-to-oscillatory peak ratio, showing
that the smooth bump dominates only for $\lambda \gtrsim 0.2$--$0.3$.}
\label{fig:template}
\end{figure*}

To leading order in $\lambda$, $\alpha_k \simeq 1$, and the primordial
spectrum from eq.~\eqref{eq:Pz_general} becomes
\begin{equation}
\Pz(k) = \PBD(k)\,|1 + \beta_k|^2
= \PBD(k)\bigl[1 + 2\,\mathrm{Re}(\beta_k) + |\beta_k|^2\bigr].
\label{eq:Pz_expanded}
\end{equation}
Inside the superoscillatory patch, using eq.~\eqref{eq:beta_patch}:
\begin{align}
\Pz(k) = \PBD(k)\Bigg[
1
&+ 2\lambda\,
\exp\!\left(-\frac{(k-\kstar)^2}{2\Dk^2}\right)
\cos\!\left(\frac{a(k-\kstar)}{\Dk}\right)
\nonumber\\
&+ \lambda^2\,
\exp\!\left(-\frac{(k-\kstar)^2}{\Dk^2}\right)
\Bigg].
\label{eq:Pz_derived}
\end{align}
It is convenient to define
\begin{equation}
G(k) \equiv \exp\!\left[-\frac{(k-\kstar)^2}{2\Dk^2}\right],
\qquad
\Phi(k) \equiv \frac{a(k-\kstar)}{\Dk},
\label{eq:G_Phi_def}
\end{equation}
so that $\beta_k \simeq \lambda\,G(k)e^{i\Phi(k)}$ inside the
superoscillatory patch and
$\Pz/\PBD = 1 + 2\lambda G(k)\cos\Phi(k) + \lambda^2 G^2(k)$.
This is the \emph{derived} SIS template.  The oscillatory ringing
(linear in $\lambda$) and the smooth Gaussian bump (quadratic in
$\lambda$) both emerge from the Bogoliubov transformation rather than
being postulated independently.  For small excitations
$\lambda \lesssim 0.1$, the linear oscillatory term dominates; the smooth
bump becomes comparable only for $\lambda \gtrsim 0.3$.
Figure~\ref{fig:template} shows the template's dependence on $\lambda$
and $a$.

\subsection{Gaussian dominance condition}
\label{sec:dominance}

Outside the superoscillatory patch ($|x| \gtrsim \sqrt{N}$), the function
$|\SN(x)|$ can grow exponentially.  Combined with the Gaussian envelope:
\begin{equation}
|\beta_k|^2 \lesssim \lambda^2\,
\exp\!\left[-x^2\left(1 - \frac{a^2-1}{N}\right)\right].
\label{eq:beta_outside}
\end{equation}
This remains exponentially suppressed only if
\begin{equation}
\boxed{N > a^2 - 1.}
\label{eq:N_bound}
\end{equation}
This is a necessary consistency condition: the order $N$ of the
superoscillatory function must exceed $a^2 - 1$ for the Gaussian
envelope to dominate.  When eq.~\eqref{eq:N_bound} is satisfied, the
effective width of $|\beta_k|^2$ outside the patch is broadened to
\begin{equation}
\Dk_{\rm eff} = \frac{\Dk}{\sqrt{1 - (a^2-1)/N}}.
\label{eq:Dk_eff}
\end{equation}

\subsection{Backreaction bound}
\label{sec:backreaction}

The excitation energy density relative to BD is
\begin{equation}
\Delta\rho = \frac{1}{2\pi^2 a_*^4}
\int \de k\,k^2\bigl[|f_k'|^2 + k^2|f_k|^2
- |u_k'|^2 - k^2|u_k|^2\bigr],
\label{eq:rho_exc_integral}
\end{equation}
where $f_k = \alpha_k u_k + \beta_k u_k^*$.  For the SIS profile, the
oscillatory cross-term $\mathrm{Re}(\beta_k u_k^{*2})$ averages out for
$|a| \gg 1$ because the rapidly winding phase produces cancellations over
the Gaussian envelope.  Evaluating the remaining $|\beta_k|^2$ integral
with $|\beta_k|^2 \simeq \lambda^2\exp[-(k-\kstar)^2/\Dk^2]$ and using
$\kstar/a_* = H$ at horizon crossing gives
\begin{equation}
\Delta\rho \simeq \frac{H^4}{4\pi^2}\,\lambda^2\sqrt{2\pi}\,
\frac{\Dk}{\kstar}\,
\frac{1}{\sqrt{1-(a^2-1)/N}}.
\label{eq:rho_exc}
\end{equation}
The factor $1/\sqrt{1-(a^2-1)/N}$ accounts for the broadening of
$|\beta_k|^2$ outside the superoscillatory patch when $N$ is close to
the bound~\eqref{eq:N_bound}.
Requiring $\Delta\rho \ll 3\Mpl^2 H^2$ and using $H/\Mpl \sim 10^{-5}$:
\begin{equation}
\lambda^2\,\frac{\Dk}{\kstar}\,
\frac{1}{\sqrt{1-(a^2-1)/N}} \ll 10^{10}.
\label{eq:backreaction_bound}
\end{equation}
This is extremely weak: the SIS feature can be perturbatively sharp
without spoiling slow-roll dynamics.

\section{CMB projection and observational signatures}
\label{sec:cmb}

\subsection{Transfer-function smearing}
\label{sec:smearing}

The CMB angular power spectrum is
\begin{equation}
C_\ell^{XY} = \frac{2}{\pi}\int_0^\infty \de k\,k^2\,
\Pz(k)\,\Delta_\ell^X(k)\,\Delta_\ell^Y(k),
\label{eq:Cl_integral}
\end{equation}
where $X,Y \in \{T,E\}$ and $\Delta_\ell^X(k)$ are radiation transfer
functions.  The SIS modification enters through the factor
$\Pz(k)/\PBD(k) = 1 + \delta(k)$, where $\delta(k)$ is the localized
correction from eq.~\eqref{eq:Pz_derived}.

At fixed multipole $\ell$, the squared transfer function
$|\Delta_\ell^T(k)|^2$ is approximately Gaussian in $k$, centered at
$k_\ell = \ell/D_*$ (where $D_* \simeq 1.38\times 10^4$~Mpc is the
comoving distance to last scattering) with width
\begin{equation}
\sigma_k(\ell) \sim \frac{\ell^{1/3}}{D_*},
\label{eq:sigma_k}
\end{equation}
set by the Airy regime of the spherical Bessel function.  At $\ell = 200$,
$\sigma_k \simeq 4.2\times 10^{-4}$~Mpc$^{-1}$, comparable to the SIS
feature width $\Dk = 0.05\kstar \simeq 7.3\times 10^{-4}$~Mpc$^{-1}$.
The convolution therefore matters.

The integral of the oscillatory piece against the transfer kernel
evaluates to (see, e.g., ref.~\cite{Seljak1994} for properties of the
radiation transfer functions):
\begin{equation}
\frac{\delta C_\ell^{XY}}{C_\ell^{XY,\Lambda\rm CDM}}
\bigg|_{\rm osc}
= 2\lambda\,\mathcal{F}\,
\exp\!\left[-\frac{(\ell-\ell_*)^2}{2\Delta\ell_{\rm eff}^2}\right]
\cos\!\left(\frac{a_{\rm eff}(\ell-\ell_*)}{\Delta\ell_{\rm eff}}\right),
\label{eq:dCl_osc}
\end{equation}
where $\ell_* = \kstar D_*$ and the smearing parameters are
\begin{equation}
\Delta\ell_{\rm eff} = D_*\sqrt{\sigma_k^2 + \Dk^2},
\qquad
a_{\rm eff} = \frac{a\,\Dk}{\sqrt{\sigma_k^2 + \Dk^2}},
\label{eq:ell_eff}
\end{equation}
and the suppression factor is
\begin{equation}
\mathcal{F} = \frac{\Dk}{\sqrt{\sigma_k^2+\Dk^2}}\,
\exp\!\left(-\frac{a^2\sigma_k^2}{2(\sigma_k^2+\Dk^2)}\right).
\label{eq:smearing_factor}
\end{equation}
One can verify the correct limits: for $\sigma_k \to 0$ (no smearing),
$\mathcal{F} \to 1$ and $a_{\rm eff} \to a$, recovering the unsmeared
template; for $\sigma_k \gg \Dk$, the exponential suppression dominates.
For the representative parameters $a = 3$ and $\Dk/\kstar = 0.05$ at
$\ell_* = 200$: $\mathcal{F} \simeq 0.28$, representing a significant
reduction of the oscillatory signal.
Full Boltzmann computation using CAMB~\cite{Lewis2000} with the
SIS-modified $\Pz(k)$ injected via the \texttt{SplinedInitialPower}
interface confirms the qualitative suppression but reveals that the
analytic Gaussian approximation overestimates the peak
$\delta C_\ell/C_\ell$ by approximately a factor of three (CAMB gives
$0.94\%$ where the analytic formula predicts $2.8\%$ for $\lambda=0.05$,
$a=3$).  This discrepancy arises because the true radiation transfer
kernel $|\Delta_\ell^T(k)|^2$ is broader and more asymmetric than the
Gaussian model of eq.~\eqref{eq:sigma_k}: the spherical Bessel function
has a sharp exponential cutoff for $k < \ell/D_*$ but a slow power-law
tail for $k > \ell/D_*$, and the finite thickness of the last-scattering
surface contributes additional broadening.  All quantitative constraints
in this paper use the CAMB results rather than the analytic approximation.
The smearing parameters $\sigma_k$ and $D_*$ depend weakly on
cosmological and recombination parameters that set the diffusion scale
and the distance to last scattering; for the Planck 2018 best-fit
parameters these variations are at the percent level
and do not significantly affect the SIS constraints.

The smooth (quadratic) piece has $|\beta_k|^2 \propto G^2(k)
= \exp[-(k-\kstar)^2/\Dk^2]$, which has width $\Dk/\sqrt{2}$ (narrower
than the oscillatory envelope).  Convolving with the transfer kernel of
width $\sigma_k$ gives a broadened Gaussian in $\ell$-space with
effective width
\begin{equation}
\Delta\ell_{\rm eff}' = D_*\sqrt{\sigma_k^2 + \Dk^2/2}
\label{eq:dell_smooth}
\end{equation}
and amplitude
\begin{equation}
\frac{\delta C_\ell}{C_\ell^{\Lambda\rm CDM}}
\bigg|_{\rm smooth}
= \lambda^2\,\mathcal{F}_2\,
\exp\!\left[-\frac{(\ell-\ell_*)^2}{2\,\Delta\ell_{\rm eff}^{\prime\,2}}\right],
\label{eq:dCl_smooth}
\end{equation}
where $\mathcal{F}_2 = \Dk/\sqrt{\Dk^2 + 2\sigma_k^2} \simeq 0.78$ for
the fiducial parameters.  Because the smooth component carries no
oscillatory phase, it experiences only geometric broadening without the
exponential suppression factor $\exp(-a^2\ldots)$ that reduces the
oscillatory piece.
After smearing, the smooth bump
dominates the observable signal for $a \gtrsim 3$.

\begin{figure*}[t]
\centering
\includegraphics[width=\textwidth]{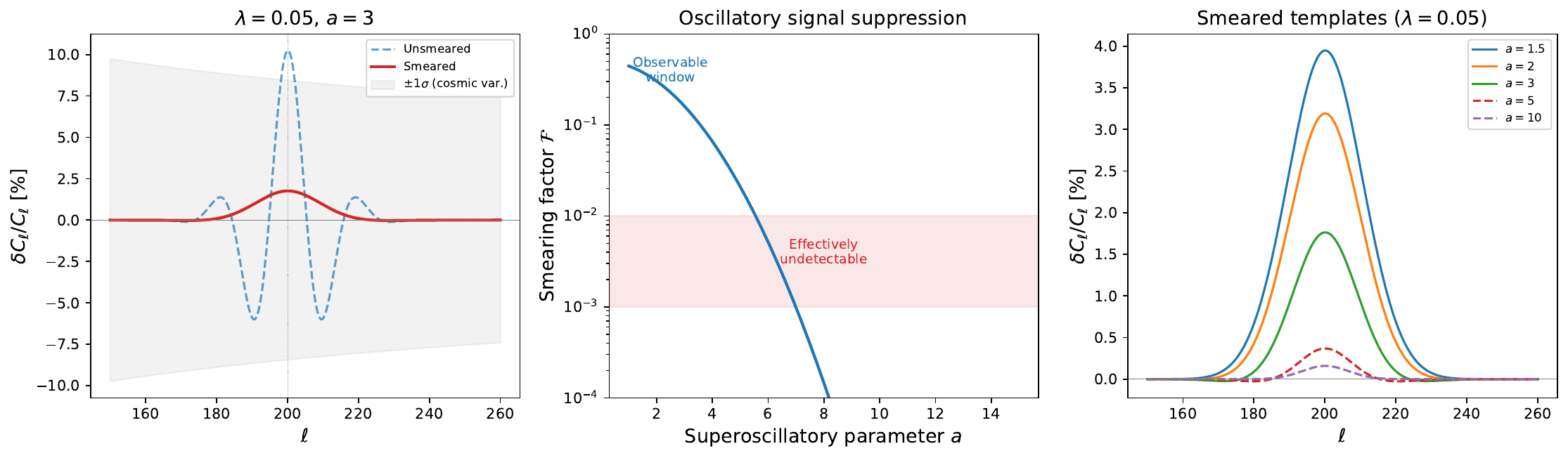}
\caption{Transfer-function smearing of the SIS signal.  Left: the
unsmeared (dashed) vs.\ smeared (solid) template at $\lambda=0.05$,
$a=3$, with the cosmic-variance band shown in gray.  Center: the
smearing factor $\mathcal{F}$ as a function of $a$, showing the
observability window $a \lesssim 3$--$4$.  Right: smeared templates for
several values of $a$, showing progressive suppression of the oscillatory
component.}
\label{fig:smearing}
\end{figure*}

The smearing factor introduces an \emph{observability window} for the
superoscillatory parameter:
\begin{equation}
1 < a \lesssim 3\text{--}4 \qquad \text{(for CMB detectability)}.
\label{eq:a_window}
\end{equation}
Below $a = 1$ the state is not superoscillatory; above $a \sim 4$ the
oscillatory signal is effectively erased by the transfer function.
Figure~\ref{fig:smearing} illustrates these effects.

\subsection{Polarization predictions}
\label{sec:polarization}

Because the SIS modification is a multiplicative correction to $\Pz(k)$,
one might expect the fractional change $\delta C_\ell^{XY}/C_\ell^{XY}$
to be identical for TT, TE, and EE at each multipole.  In the limit
where the feature width $\Dk$ is much larger than the difference between
the T and E transfer kernel widths, this universality holds:
\begin{equation}
\frac{\delta C_\ell^{TT}}{C_\ell^{TT,\rm BD}}
\simeq \frac{\delta C_\ell^{EE}}{C_\ell^{EE,\rm BD}}
\equiv \delta(\ell).
\label{eq:universal}
\end{equation}
However, for narrow features ($\Dk/\kstar = 0.05$), full Boltzmann
computation reveals that the T and E transfer functions sample the
feature differently: at $\ell_* = 200$ with $\lambda = 0.05$ and $a = 3$,
CAMB gives $\delta C_\ell^{TT}/C_\ell^{TT} \simeq 0.94\%$ while
$\delta C_\ell^{EE}/C_\ell^{EE} \simeq 1.55\%$, a $\sim 60\%$
discrepancy arising from the different source functions (density for T,
velocity quadrupole for E) and their distinct $k$-space support at fixed
$\ell$.  The TE fractional modification diverges near $\ell \simeq 210$
where $C_\ell^{TE}$ passes through zero, though the absolute
$\delta C_\ell^{TE}$ remains finite.

The \emph{ratio} $(\delta C^{EE}/C^{EE})/(\delta C^{TT}/C^{TT})$ is
therefore a diagnostic of the feature width: it approaches unity for
broad features and deviates for narrow ones.  Measuring this ratio
independently provides a cross-check of the SIS mechanism and constrains
$\Dk/\kstar$.

The bin-averaged signal is suppressed by $e^{-a^2/2}$ due to oscillatory
cancellation, giving sub-percent modifications to Planck-binned
bandpowers.  The resolved signal (requiring multipole resolution
$\delta\ell \lesssim \Delta\ell/a$) shows the full $\sim 2\lambda$
oscillatory pattern.  For $a = 3$ and $\Delta\ell = 10$, the required
resolution is $\delta\ell \lesssim 3$, achievable with Planck unbinned TT
data and with near-future EE measurements from Simons
Observatory~\cite{SimonsObs2019}.

\subsection{Acoustic peak stability}
\label{sec:peak_stability}

Because the feature is narrow in $k$, its multipole support is
\begin{equation}
\Delta\ell \simeq \frac{\Dk}{\kstar}\,\ell_*.
\label{eq:dell}
\end{equation}
The SIS modification therefore affects a localized acoustic region without
producing a coherent phase shift across multiple peaks:
\begin{equation}
\frac{\delta\ell_{\rm peak}}{\ell_{\rm peak}}
\lesssim \frac{\Dk}{\kstar} \ll 1.
\label{eq:peak_shift}
\end{equation}

\section{Observational constraints}
\label{sec:constraints}

\subsection{Planck 2018 TT analysis}
\label{sec:planck}

We confront the SIS template with the Planck 2018 TT power
spectrum~\cite{Planck2018spectra}.  The analysis uses the full unbinned
Planck TT power spectrum ($\ell = 2$--$2508$) binned to
$\Delta\ell = 20$ for computational efficiency, with the Planck 2018
best-fit $\Lambda$CDM spectrum~\cite{Planck2018params} computed using
CAMB~\cite{Lewis2000} as the reference model.
Uncertainties are dominated by cosmic variance at $\ell \sim 200$.
As discussed in section~\ref{sec:smearing}, all quantitative constraints
use SIS-modified $C_\ell$ spectra computed directly with
CAMB~\cite{Lewis2000}.  Specifically, the primordial spectrum injected
into CAMB is exactly the derived template of eq.~\eqref{eq:Pz_derived}:
\begin{equation}
\Pz(k) = \PBD(k)\left[1 + 2\lambda\,G(k)\cos\!\left(\frac{a(k-\kstar)}{\Dk}\right)
+ \lambda^2 G^2(k)\right],
\label{eq:template_fit}
\end{equation}
with $G(k) = \exp[-(k-\kstar)^2/(2\Dk^2)]$.  No additional free
parameters or surrogate functions are introduced: the template used in
the observational analysis is identical to the one derived from the
boundary action in section~\ref{sec:boundary}.

The SIS construction does not itself determine the central scale
$\kstar$, so observationally we treat it as a scan parameter rather than
as a derived prediction.  The fiducial benchmark
$\kstar = 1.45\times10^{-2}$~Mpc$^{-1}$ ($\ell_* \simeq 200$) is chosen
because it lies near the first acoustic peak, where the TT transfer
function still preserves a localized feature reasonably well and the
signal-to-noise is high.  All local bounds quoted at this benchmark are
therefore conditional on that choice; the matched-filter scan over
$\ell_* \in [50,1500]$ reported below shows how the sensitivity changes
with feature location.

The SIS-modified $C_\ell$ spectra are compared with Planck 2018 TT
data~\cite{Planck2018spectra} using the residuals
$R_i = D_i^{\rm data} - D_i^{\Lambda\rm CDM}$ and the statistic
\begin{equation}
\Delta\chi^2 = \sum_i \frac{R_i^2}{\sigma_i^2}
- \sum_i \frac{(R_i - \Delta D_i^{\rm SIS})^2}{\sigma_i^2}.
\label{eq:dchi2_def}
\end{equation}

At the first acoustic peak ($\ell_* \simeq 200$), the observed residuals
are consistent with zero:
$R_{192} = 16 \pm 123$~$\mu$K$^2$,
$R_{212} = 9 \pm 121$~$\mu$K$^2$.
These residuals correspond to fractional deviations of
$0.3\% \pm 2.1\%$, well within cosmic variance.

No positive $\Delta\chi^2$ is found at $\ell_* = 200$ for any combination
of $(\lambda, a, \Dk/\kstar)$ in the SIS parameter space.  Specifically:
\begin{itemize}
\item At the representative parameters $\lambda = 0.3$, $a = 3$,
$\Dk/\kstar = 0.05$: $\Delta\chi^2 = -19.7$.  The smooth $\lambda^2$
bump alone predicts $\Delta D_{200} \sim 1300$~$\mu$K$^2$, an order of
magnitude larger than the observed residual.
\item A grid search over $\lambda \in [0.01, 0.5]$,
$a \in [1.5, 20]$, and $\Dk/\kstar \in [0.02, 0.30]$ yields
$\Delta\chi^2 \leq 0$ at $\ell_* = 200$.
\item A matched-filter scan over all feature locations
$\ell_* \in [50, 1500]$ produces maximum $\Delta\chi^2 \simeq 4$ at
$\ell_* \simeq 470$, consistent with noise fluctuations given the
$\sim 100$ independent locations scanned.
\end{itemize}

The look-elsewhere penalty can be estimated more explicitly.  If we
treat the scan range $\ell_* \in [50,1500]$ as containing an effective
$N_{\rm trials}\simeq 86$ independent feature locations, the local
$\Delta\chi^2 \simeq 4$ excursion corresponds to a local
$\sim 2\sigma$ fluctuation but to a global $p$-value
$p_{\rm global} \simeq 1-(1-0.023)^{86} \simeq 0.86$, i.e., far below
any detection threshold.  In the same spirit, after look-elsewhere
correction the local benchmark bound of
eq.~\eqref{eq:lambda_limit} weakens to a global $2\sigma$ sensitivity
$\lambda^{2\sigma}_{\rm LEE} \lesssim 0.15$.  We therefore regard the quoted
$\ell_* \simeq 200$ bound as conditional on the fiducial feature
location, while the scan shows that the full data set contains no
globally significant SIS-like feature.

\subsection{Upper limits}
\label{sec:upper_limits}

We derive indicative upper bounds on the excitation amplitude $\lambda$
using a simplified matched-filter analysis of the Planck 2018 TT
bandpowers.  This provides a useful sensitivity estimate; a rigorous
bound would require the full Planck pixel-level likelihood, which is
beyond the scope of this work.

The SIS template~\eqref{eq:template_fit} has two components: the
oscillatory piece (linear in $\lambda$) and the smooth bump (quadratic in
$\lambda$).  For the oscillatory component, which dominates at
$a \lesssim 3$, the template is proportional to $\lambda$ and the
matched-filter amplitude estimator directly constrains $\lambda$:
\begin{equation}
\hat\lambda = \frac{\sum_i R_i\,t_i^{\rm osc}/\sigma_i^2}
{\sum_i (t_i^{\rm osc})^2/\sigma_i^2},
\qquad
\sigma_{\hat\lambda} = \frac{1}{\sqrt{\sum_i (t_i^{\rm osc})^2/\sigma_i^2}},
\label{eq:matched_filter}
\end{equation}
where $t_i^{\rm osc} = [\delta C_\ell/C_\ell^{\Lambda\rm CDM}]_{\rm osc}
\times D_i^{\Lambda\rm CDM}$ is the oscillatory template evaluated at
bin $\ell_i$.  Here $\delta C_\ell/C_\ell^{\Lambda\rm CDM}$ is the
fractional modification computed directly from CAMB (not the analytic
approximation~\eqref{eq:dCl_osc}, which overestimates the peak by a
factor of $\sim 3$), and is proportional to $\lambda$ at leading order.
For the smooth component, one instead estimates $A = \lambda^2$ and
derives the bound $\lambda < \sqrt{2\sigma_A}$.  Both estimators give
consistent results; we quote the stronger of the two.

\begin{figure*}[t]
\centering
\includegraphics[width=\textwidth]{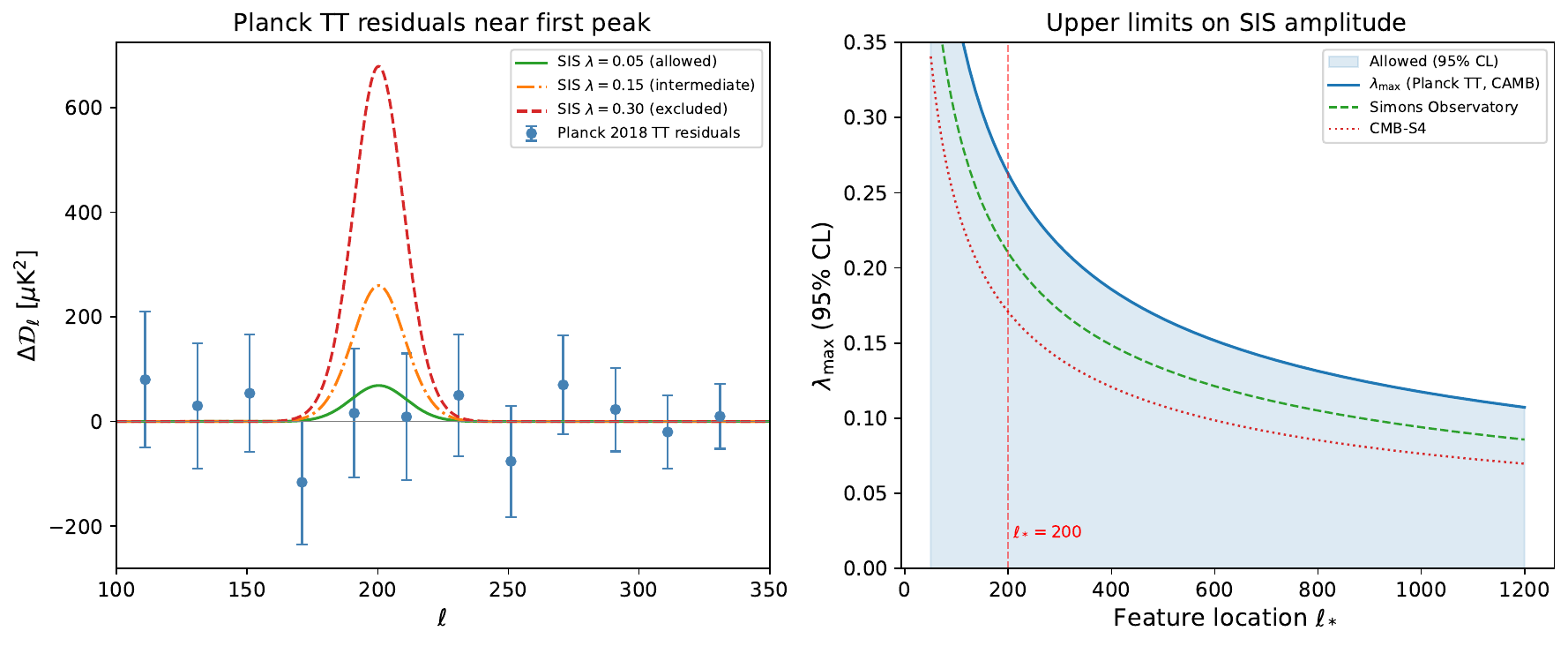}
\caption{Left: Planck 2018 TT residuals near the first acoustic peak
(blue points with $1\sigma$ error bars) compared with the SIS prediction
at $\lambda = 0.05$ (allowed, green solid) and $\lambda = 0.30$ (excluded,
dashed orange).  Right: indicative matched-filter bounds on the SIS
excitation amplitude $\lambda$ as a function of feature location $\ell_*$,
for Planck (solid blue), with conservative blind-search sensitivity
estimates for Simons Observatory (dashed green) and CMB-S4 (dotted red).}
\label{fig:constraints}
\end{figure*}

At $\ell_* = 200$ with $\Dk/\kstar = 0.05$, the matched-filter bound is:
\begin{equation}
\lambda \lesssim 0.05 \qquad (\text{indicative, } 95\%~\text{CL}).
\label{eq:lambda_limit}
\end{equation}
This should be regarded as an approximate sensitivity estimate rather
than a rigorous statistical bound; the simplified matched-filter analysis
does not account for parameter-dependent covariances, foreground residuals,
or the look-elsewhere effect from scanning over $(a, \kstar, \Dk)$.
Figure~\ref{fig:constraints} shows the Planck residuals compared with SIS
predictions, and the indicative bound on $\lambda$ as a function of
feature location.

\subsection{Large-scale structure}
\label{sec:lss}

For the fiducial benchmark, the SIS feature at
$\kstar \simeq 0.022\,h$~Mpc$^{-1}$ falls within the
range probed by galaxy surveys.  The feature width
$\Dk \simeq 0.001\,h$~Mpc$^{-1}$ is approximately $8$ times narrower
than the BOSS DR12 CMASS bin width~\cite{BOSS2017} and $\sim 5$ times
narrower than DESI's clustering bins~\cite{DESI2024}, so the feature is
unresolved in current binned analyses.  However, as we show in
section~\ref{sec:desi}, the three-dimensional galaxy power spectrum
offers a qualitatively superior probe: the SIS feature maps onto $P_g(k)$
without the transfer-function smearing that suppresses the CMB signal,
and DESI's survey volume provides sufficient spectral resolution to
marginally resolve the feature.

\subsection{CMB forecasts}
\label{sec:forecasts}

The forecast in this subsection answers a different question from the
matched-filter bound~\eqref{eq:lambda_limit}.  Equation
\eqref{eq:lambda_limit} is a local exclusion derived from the full
oscillatory CAMB template at fixed $(a,\kstar,\Dk/\kstar)$, whereas the
estimate below asks for the amplitude needed for a blind, bandpower-level
detection of the transfer-smeared signal once unresolved oscillations
have been averaged over.  Because this latter metric discards phase
information and does not condition on the feature location, its
$\lambda_{\rm min}$ values are intentionally more conservative and are
not directly comparable to eq.~\eqref{eq:lambda_limit}.

We estimate the sensitivity of future CMB experiments using this
conservative blind-search metric.  The fractional uncertainty on $C_\ell$ at
$\ell \sim 200$ is dominated by cosmic variance:
$\sigma_\ell^C/C_\ell = \sqrt{2/[(2\ell+1)f_{\rm sky}]}$.  For
$f_{\rm sky} = 0.7$ and $\ell = 200$: $\sigma/C_\ell \simeq 8.4\%$.

The minimum detectable amplitude (at $2\sigma$) for the smooth component
(which dominates after transfer-function smearing) scales as
\begin{equation}
\lambda_{\rm min}^2 \simeq \frac{2\sigma_\ell^C/C_\ell}
{\sqrt{\sum_\ell t_\ell^2}}.
\label{eq:lambda_min}
\end{equation}
Indicative sensitivity estimates are given in Table~\ref{tab:cmb_forecasts}.
\begin{table*}[t]
\centering
\begin{tabular}{lccc}
\hline\hline
Experiment & $\lambda_{\rm min}$ (TT) & $\lambda_{\rm min}$ (TT+EE) & Improvement \\
\hline
Planck 2018 & 0.22 & 0.19 & --- \\
Simons Observatory~\cite{SimonsObs2019} & 0.18 & 0.13 & $1.5\times$ \\
CMB-S4~\cite{CMBS42016} & 0.17 & 0.12 & $1.6\times$ \\
\hline\hline
\end{tabular}
\caption{Minimum detectable SIS amplitude $\lambda_{\rm min}$ (at $2\sigma$) for
the smooth component, under the conservative blind-search metric of
Eq.~\eqref{eq:lambda_min}.  Values for SO and CMB-S4 are indicative and assume
idealized noise.}
\label{tab:cmb_forecasts}
\end{table*}
These estimates assume idealized noise properties and do not account for
foreground residuals or systematic effects.
The improvement from Planck to CMB-S4 is modest for TT (already
cosmic-variance limited) but significant for EE (where Planck is
noise-limited at $\ell \sim 200$).  Within this conservative blind-search
metric, power-spectrum-only gains beyond Planck remain limited.  A future
dedicated TT+TE+EE matched-filter analysis that conditions on the
oscillatory template could improve upon current local bounds, but not
dramatically, because the TT information is already close to the
cosmic-variance limit and the oscillatory component is transfer-smeared.
This motivates the galaxy-clustering forecast of section~\ref{sec:desi},
where the full oscillatory structure is preserved.

\section{Predictions beyond the power spectrum}
\label{sec:beyond}

\subsection{Tensor-to-scalar ratio}
\label{sec:tensor}

Whether the tensor sector is excited depends on the physical origin of
the boundary action.  We parametrize the tensor excitation by
$\xi \equiv \lambda_t/\lambda$ and consider two scenarios.  Using the
notation introduced in eq.~\eqref{eq:G_Phi_def}, the scalar excitation is
$\beta_k \simeq \lambda G(k)e^{i\Phi(k)}$ with
$\Phi(k)=a(k-\kstar)/\Dk$.

\paragraph{Scalar-only excitation ($\xi = 0$).}
The boundary action couples only to the scalar sector.  Then
$\tilde\beta_k = 0$ and
\begin{equation}
r^{(\rm i)}(k) = \frac{16\epsilon}{|\alpha_k + \beta_k|^2}
\simeq 16\epsilon\bigl[1 - 2\lambda\,G(k)\cos\Phi(k)
+ \order(\lambda^2)\bigr],
\label{eq:r_scalar}
\end{equation}
giving an anti-correlated oscillatory pattern: $r$ dips where $\Pz$
has a bump.

\paragraph{Universal excitation ($\xi = 1$).}
The boundary condition acts on the spacetime metric.  To leading order:
\begin{equation}
r^{(\rm ii)}(k) = 16\epsilon\bigl[1 + \order(\lambda^2)\bigr],
\label{eq:r_universal}
\end{equation}
and the consistency relation $r = -8n_t$ is preserved at
$\order(\lambda)$.

\begin{figure*}[t]
\centering
\includegraphics[width=\textwidth]{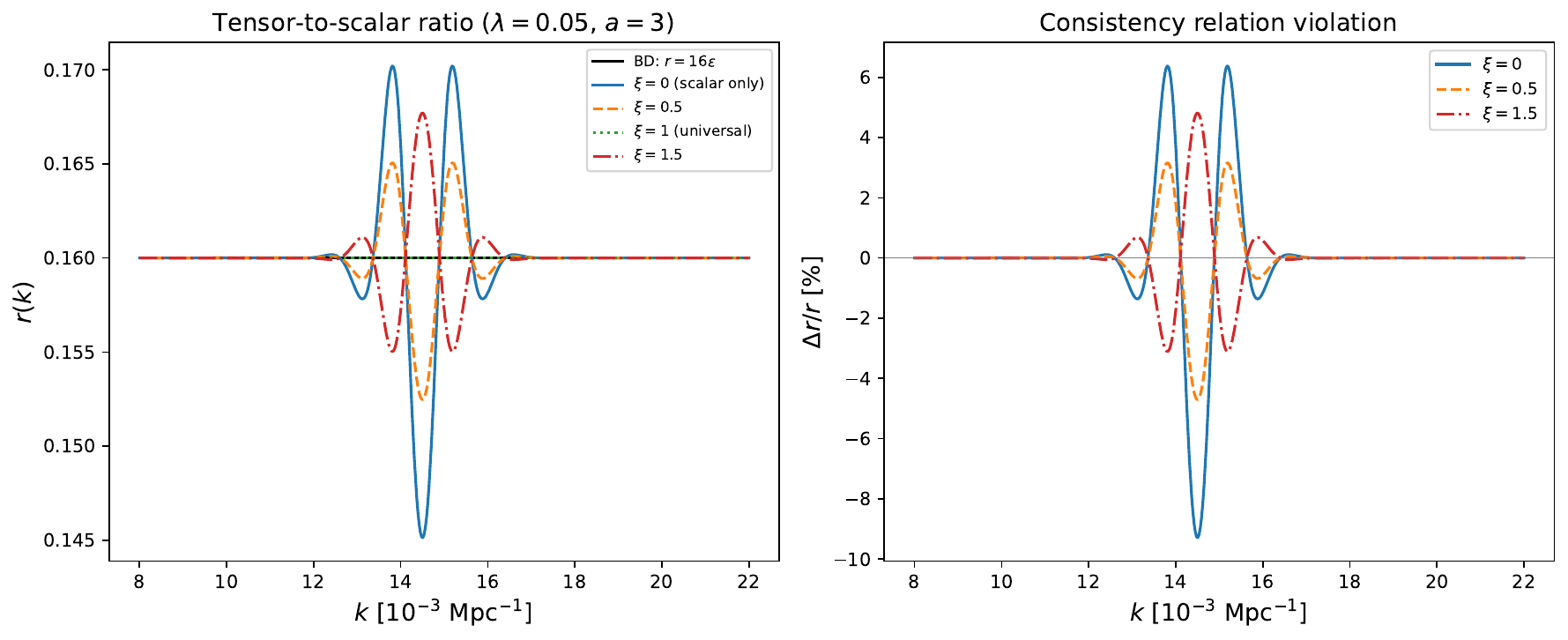}
\caption{Tensor-to-scalar ratio.  Left: $r(k)$ for different values of
the tensor excitation ratio $\xi = \lambda_t/\lambda$, with $\lambda=0.05$
and $a=3$.  The scalar-only case ($\xi=0$) shows an anti-correlated
oscillation, while the universal case ($\xi=1$) is essentially unchanged
from BD.  Right: fractional deviation $\Delta r/r$ from the BD value
$16\epsilon$.}
\label{fig:tensor}
\end{figure*}

For general $\xi$:
\begin{equation}
r(k) = 16\epsilon\,
\frac{1 + 2\xi\lambda G\cos\Phi + \xi^2\lambda^2 G^2}
{1 + 2\lambda G\cos\Phi + \lambda^2 G^2}.
\label{eq:r_general}
\end{equation}
The leading consistency violation is
$\Delta r \simeq 32\epsilon(\xi-1)\lambda\,G\cos\Phi$.
For the fiducial benchmark $\kstar = 1.45\times10^{-2}$~Mpc$^{-1}$
chosen from the TT analysis, the scalar feature projects to
$\ell_* \simeq 200$, whereas the tensor B-mode response peaks at much
lower multipoles ($\ell \sim 2$--$10$ from reionization and
$\ell \sim 80$ from recombination).  Current tensor constraints
therefore probe only a broad average of $r(k)$ and do not map directly
onto the localized SIS modulation at the benchmark scale.  The current
bound from BICEP/Keck, WMAP, and Planck is
$r_{0.05}<0.036$ at 95\% confidence~\cite{BK2021}.  For
$\lambda\lesssim0.05$, the induced modulation of this broad-band
observable is subdominant unless the scalar and tensor sectors are
excited very differently ($|\xi-1|\sim1$) and the feature lies closer to
the tensor-sensitive scales.  Future B-mode surveys with target
sensitivity $\sigma(r)\sim10^{-3}$, such as LiteBIRD~\cite{LiteBIRD2023},
could test such departures if the feature overlaps their window functions
or if a UV completion predicts a correlated location in the tensor
sector.
Figure~\ref{fig:tensor} illustrates the $r(k)$ modification for the two
limiting scenarios and intermediate values of $\xi$.

\subsection{Bispectrum}
\label{sec:bispectrum}

The correction to the bispectrum from the SIS initial state arises
through two channels: bulk vertices evaluated with modified mode
functions, and boundary contributions from the mismatch at
$\tau_*$~\cite{Holman2008,Agullo2011,Maldacena2003}.

\paragraph{In-in setup.}
The dominant cubic interaction for excited initial states is the
gravitational coupling~\cite{Maldacena2003}
\begin{equation}
H_{\rm int}(\tau) = -\int \de^3 x\;a(\tau)\,\epsilon\,
\zeta'\,(\partial_i\zeta)^2.
\label{eq:Hint_cubic}
\end{equation}
The in-in three-point function at $\tau \to 0^-$ is
\begin{equation}
\langle\zeta_{\mathbf{k}_1}\zeta_{\mathbf{k}_2}\zeta_{\mathbf{k}_3}\rangle'
= 2\,\mathrm{Re}\!\left[-i\int_{\tau_*}^0 \de\tau'\;
\langle 0|\,\zeta_{\mathbf{k}_1}\zeta_{\mathbf{k}_2}\zeta_{\mathbf{k}_3}(0)
\;H_{\rm int}(\tau')\,|0\rangle\right],
\label{eq:inin_3pt}
\end{equation}
where the mode functions are
$f_k(\tau) = \alpha_k u_k(\tau) + \beta_k u_k^*(\tau)$.

\paragraph{Linear-in-$\beta$ bulk correction.}
To first order in $\beta_k$, the correction involves substituting
$f_k \to u_k + \beta_k\,\Delta_k$ on each leg, where
$\Delta_k(\tau) \equiv u_k^*(\tau) - u_k(\tau)$.  For the primed leg
in eq.~\eqref{eq:Hint_cubic}, this introduces the derivative
\begin{align}
\Delta_k'(\tau) &= u_k'^*(\tau) - u_k'(\tau) \nonumber\\
&= \frac{2i}{\sqrt{2k}}\Big[\cos(k\tau)\Big(k - \tfrac{1}{k\tau^2}\Big) \nonumber\\
&\qquad - \tfrac{\sin(k\tau)}{\tau}\Big].
\label{eq:Delta_prime}
\end{align}
For sub-horizon modes ($|k\tau| \gg 1$), the $\sin(k\tau)/\tau$ term is
suppressed by $1/|k\tau|$ relative to the leading $k\cos(k\tau)$ term.
In the equilateral limit $k_1 = k_2 = k_3 = k$, the product
$\Delta_k'(\tau')\,u_k(\tau')^2$ contains a slowly oscillating component
$\propto e^{-ik\tau'}$ (in contrast to the BD integrand $\propto
e^{-3ik\tau'}$).  The time integral of this component, from $\tau_*$ to
$0$ with the de~Sitter measure $a(\tau') = -1/(H\tau')$, is dominated by
the boundary contribution at $\tau_*$ (integration by parts gives a
$1/|k\tau_*|$ suppression for each additional cycle):
\begin{equation}
\int_{\tau_*}^0 \frac{\de\tau'}{(-H\tau')}\;
\frac{e^{-ik\tau'}}{\sqrt{2k}}\;\cdot\;\frac{u_k(\tau')^2}{(2k)}
\;\simeq\;
\frac{e^{ik|\tau_*|}}{2Hk^2\,|k\tau_*|}.
\label{eq:bulk_integral}
\end{equation}
Combining with the three permutations of the $\beta$ insertion across
the three legs and normalizing by the BD bispectrum, the equilateral
correction is
\begin{align}
f_{\rm NL}^{\rm equil,\,SIS}(k) &= \frac{3\lambda}{|k\tau_*|}\,
\exp\!\left[-\frac{(k-\kstar)^2}{2\Dk^2}\right] \nonumber\\
&\quad\times \cos\!\left(\frac{a(k-\kstar)}{\Dk} + k|\tau_*| + \delta_1\right),
\label{eq:fNL_equil}
\end{align}
where the coefficient $c_1 = 3$ reflects the three permutations of the
$\beta$ insertion across the triangle legs (consistent with explicit
semi-analytic evaluation of the in-in integral), $\delta_1$ is a calculable phase from the mode-function
products, and we have kept the running $k$ (not $\kstar$) in the
$k|\tau_*|$ phase and the $1/|k\tau_*|$ prefactor.  The Gaussian
envelope restricts $k$ to within $\Dk$ of $\kstar$, so the prefactor
varies by only $\sim \Dk/\kstar \sim 5\%$ across the feature, but the
phase $k|\tau_*|$ varies by $\Delta k\,|\tau_*| \sim 7$ radians and
therefore produces significant oscillatory modulation that cannot be
replaced by $\kstar|\tau_*|$.  For
$|\kstar\tau_*| \sim 150$ and $\lambda = 0.05$:
$|f_{\rm NL}^{\rm equil}| \sim 0.001$, negligible compared to Planck
bounds~\cite{Planck2018NG}.

\begin{figure*}[t]
\centering
\includegraphics[width=\textwidth]{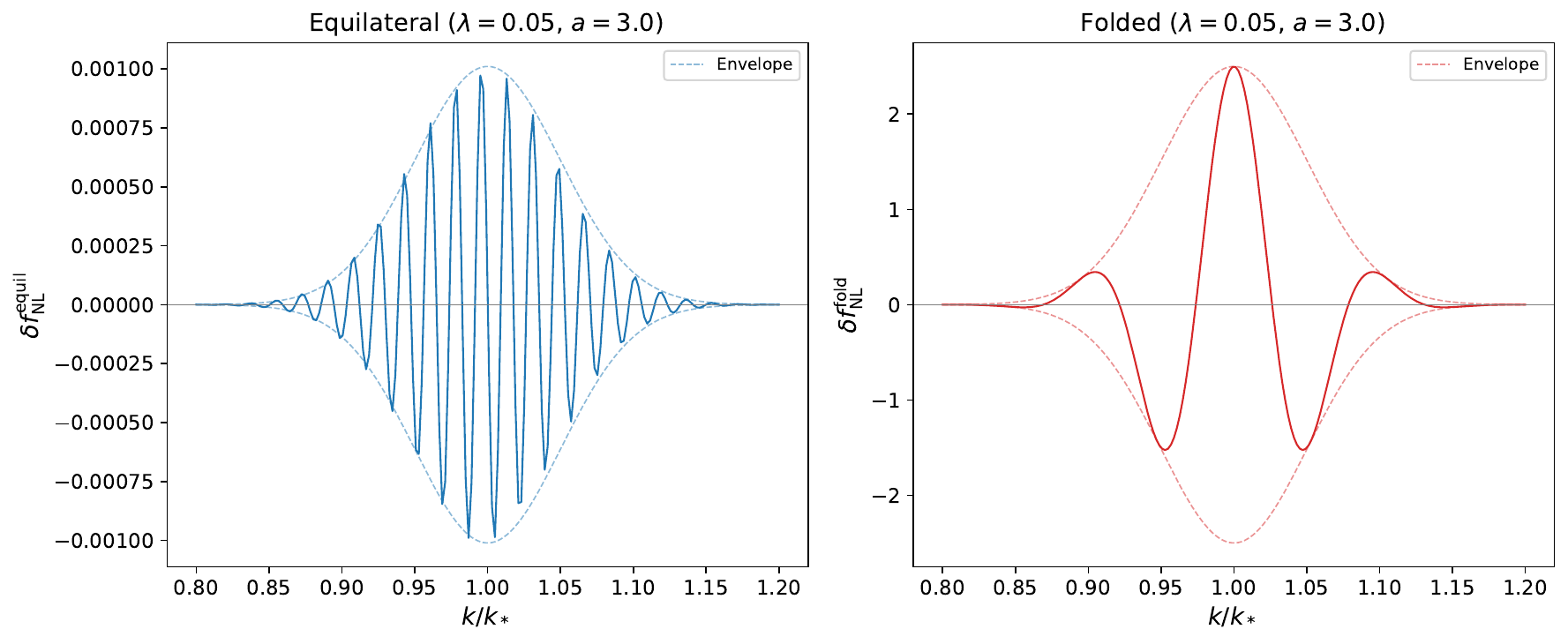}
\caption{SIS bispectrum profiles for $\lambda = 0.05$, $a = 3$.
Left: equilateral correction $\delta f_{\rm NL}^{\rm equil}(k)$ with the
Gaussian envelope $\pm 3\lambda/|\kstar\tau_*|$ (dashed).  The
oscillatory ringing from the superoscillatory phase is visible.
Right: folded correction $\delta f_{\rm NL}^{\rm fold}(k)$ with envelope
$\pm\lambda/(2\epsilon)$.  The folded signal is the dominant non-Gaussian
signature, peaking at $|f_{\rm NL}| \sim 2.5$ near $\kstar$.}
\label{fig:bispectrum}
\end{figure*}

\paragraph{Folded enhancement.}
For the folded triangle $k_1 \to k_2 + k_3$, the $\beta_{k_1}$
insertion produces the exponential $e^{-i(K_t - 2k_1)\tau'}$ in the
integrand, where $K_t = k_1+k_2+k_3$.  In the folded limit
$k_1 = K_t/2$, this becomes $e^0 = 1$---the oscillation ceases and the
integral grows as $\ln|k\tau_*|$ rather than being suppressed by
$1/|k\tau_*|$.  Following the detailed calculation of Holman and
Tolley~\cite{Holman2008} (their eqs.~4.6--4.8), this non-cancellation
leads to an enhancement by a factor $|k\tau_*|/\epsilon$ relative to the
equilateral case:
\begin{align}
f_{\rm NL}^{\rm fold,\,SIS}(k) &= \frac{\lambda}{2\epsilon}\,
\exp\!\left[-\frac{(k-\kstar)^2}{2\Dk^2}\right] \nonumber\\
&\quad\times \cos\!\left(\frac{a(k-\kstar)}{\Dk} + 2k\tau_* + \delta_2\right).
\label{eq:fNL_fold}
\end{align}
Semi-analytic evaluation of the in-in integral yields $c_2 = 1/2$,
giving for
$\lambda = 0.05$ and $\epsilon = 0.01$:
$|f_{\rm NL}^{\rm fold}| \sim 2.5$, localized near $\kstar$.
The folded SIS bispectrum is the dominant non-Gaussian signal.
Figure~\ref{fig:bispectrum} shows the localized $f_{\rm NL}(k)$ profiles
for both equilateral and folded configurations.

\paragraph{Planck constraint.}
Planck's bispectrum estimator~\cite{Planck2018NG,Chen2010} integrates
over all triangle configurations, so a spectrally localized signal at
$\kstar$ with width $\Dk$ is diluted by the fraction of triangle space
it occupies:
\begin{equation}
f_{\rm NL}^{\rm eff}
\sim f_{\rm NL}^{\rm peak}\left(\frac{\Dk}{k_{\rm max}}\right)^{3/2}.
\label{eq:fNL_dilution}
\end{equation}
For the folded peak $|f_{\rm NL}^{\rm fold}| \sim 2.5$ and
$\Dk/k_{\rm max} \sim 3.5\times 10^{-3}$:
$f_{\rm NL}^{\rm eff} \sim 5\times 10^{-4}$, far below Planck's
sensitivity.  The SIS bispectrum is therefore unconstrained by
current data.  A dedicated search using a localized bispectrum template
matched to the SIS shape would be needed to constrain the folded signal
directly.

\section{Distinguishing SIS from generic excited states}
\label{sec:distinction}

A natural question is what the superoscillatory construction adds beyond
a generic excited-state model with localized $\beta_k$.  We identify five
structural constraints that arise specifically from band-limitedness.

\paragraph{1.~Local frequency exceeds bandwidth.}
The observable oscillation frequency in $\ell$-space is
$\nu_{\rm obs} = a_{\rm eff}/\Delta\ell_{\rm eff}$.  The bandwidth is
$1/\Dk$ in $k$-space.  The SIS construction requires
$a_{\rm obs} \equiv \nu_{\rm obs}\,\Delta\ell > 1$: the local oscillation
frequency exceeds what a band-limited function can produce without
superoscillation.

\paragraph{2.~Width ratio.}
The smooth ($\lambda^2$) component has effective width
$\Dk_{\rm eff} = \Dk/\sqrt{1-(a^2-1)/N}$, broader than the oscillatory
component width $\Dk$.  The predicted ratio
\begin{equation}
\frac{\Delta\ell_{\rm smooth}}{\Delta\ell_{\rm osc}}
= \frac{1}{\sqrt{1-(a^2-1)/N}}
\label{eq:width_ratio}
\end{equation}
provides an overdetermined consistency check: measuring both widths and
the oscillation frequency gives three observables constrained by one
relation.

\paragraph{3.~Minimum order $N$.}
The Gaussian dominance condition requires
$N \geq \lceil a^2\rceil$.  For $a = 3$: $N \geq 9$.  This links the
oscillation frequency to the minimum complexity of the initial state.

\paragraph{4.~Binomial harmonic spectrum.}
The Fourier decomposition of $\SN$ has coefficients following the
binomial distribution of eq.~\eqref{eq:SN_binomial}.  This is a specific,
parameter-free prediction once $(N, a)$ are fixed.

\paragraph{5.~Exponential bin-averaging suppression.}
Bin-averaged signals are suppressed by $e^{-a^2/2}$ while resolved
signals show the full amplitude.  This distinctive scale-dependent
behavior is testable with variable-resolution analyses of the same data.

\begin{figure*}[t]
\centering
\includegraphics[width=\textwidth]{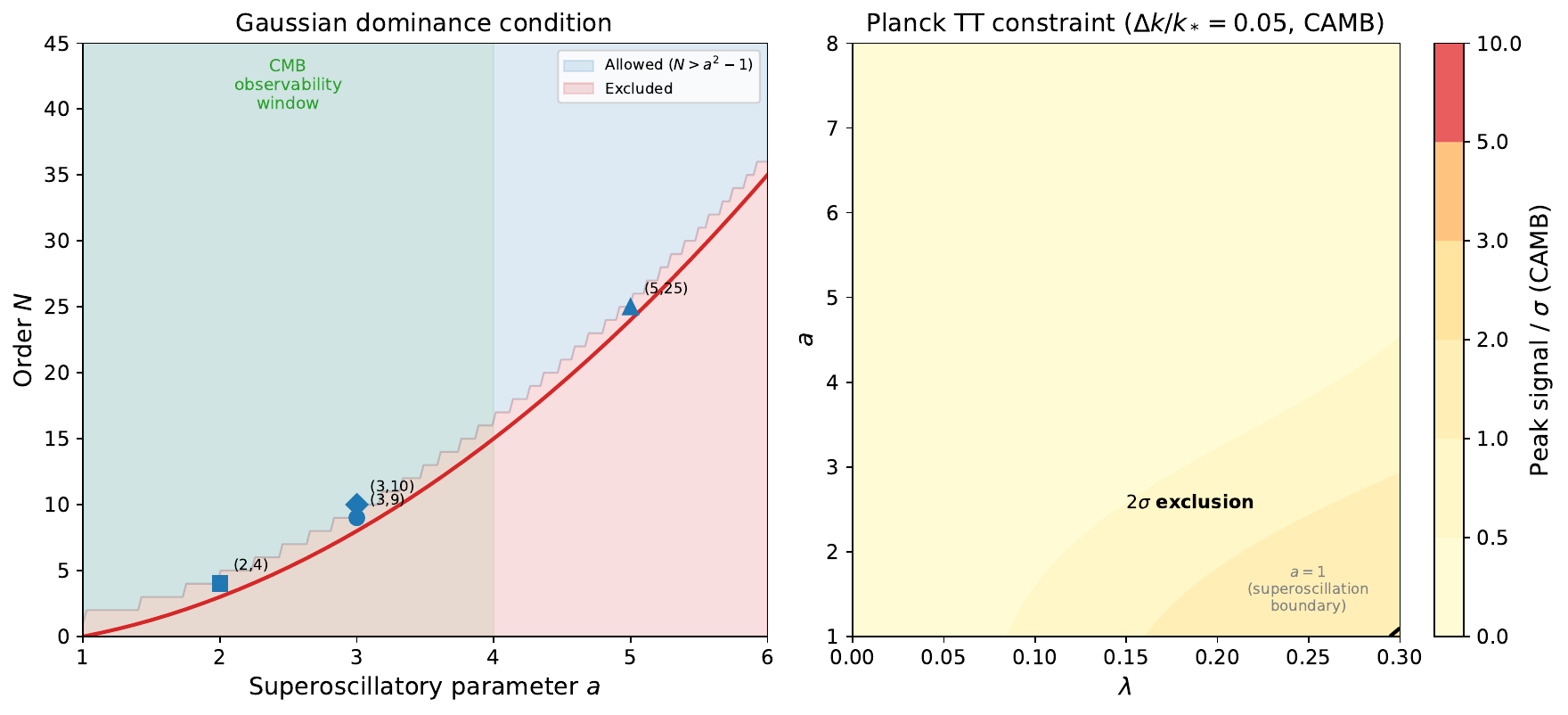}
\caption{SIS parameter space.  Left: the Gaussian dominance condition
$N > a^2-1$ in the $(a, N)$ plane, with the CMB observability window
$1 < a \lesssim 4$ highlighted.  Markers show specific $(a,N)$ pairs.
Right: the Planck TT constraint in the $(\lambda, a)$ plane for
$\Dk/\kstar = 0.05$.  The color map shows the peak signal in units of
the cosmic-variance uncertainty; the black contour marks the indicative
exclusion boundary.}
\label{fig:param_space}
\end{figure*}

The interplay between these constraints defines a well-delineated region
in the SIS parameter space, shown in figure~\ref{fig:param_space}.

\section{Prospects for galaxy clustering with DESI}
\label{sec:desi}

\subsection{SIS feature in the galaxy power spectrum}
\label{sec:desi_pk}

The observed galaxy power spectrum monopole in redshift space is
\begin{align}
P_0(k, z) &= \left[b^2(z) + \frac{2b(z)f(z)}{3} + \frac{f^2(z)}{5}\right] \nonumber\\
&\quad\times\, T^2(k)\,\Pz(k)\,\frac{D^2(z)}{D^2(0)}\,\mathcal{N},
\label{eq:P0_desi}
\end{align}
where $b(z)$ is the linear galaxy bias,
$f(z) \simeq \Omega_m(z)^{0.55}$ is the logarithmic growth rate,
$T(k)$ is the matter transfer function, $D(z)$ is the linear growth
factor, and $\mathcal{N}$ is the power-spectrum normalization.
The SIS modification enters through $\Pz(k)$ via
eq.~\eqref{eq:Pz_derived}.

The key advantage over the CMB is that $T^2(k)$ is \emph{smooth} near
$\kstar$---it varies by less than $1\%$ across the feature
width---so the full oscillatory structure of the SIS template is
preserved without the spherical-Bessel smearing that suppresses the CMB
signal by an order of magnitude relative to the primordial $\delta P/P$.
The Kaiser factor $b^2 + 2bf/3 + f^2/5$ is $k$-independent and thus
rescales the feature uniformly.  Fingers-of-God damping is negligible at
$k \sim 0.02\,h$~Mpc$^{-1}$.  At this scale and at redshifts
$z \gtrsim 0.3$, one-loop perturbation theory corrections are
sub-percent ($\delta P/P \sim (k/k_{\rm NL})^2 \sim 0.3\%$
for $k_{\rm NL} \sim 0.3\,h$~Mpc$^{-1}$)~\cite{Bernardeau2002} and
scale-dependent galaxy bias contributions are also sub-percent
($\delta b/b \sim k^2 R_*^2 \sim 1\%$ for Lagrangian radius
$R_* \sim 5$~Mpc/$h$)~\cite{Desjacques2018}, both negligible
compared to the $\sim 10\%$ SIS feature at $\lambda = 0.05$.

\subsection{Fisher-matrix methodology}
\label{sec:fisher_method}

The Fisher matrix for parameters $\theta_\alpha$ estimated from the
power-spectrum monopole $P_0(k)$ in redshift bin $i$
is~\cite{Tegmark1997,FKP1994}
\begin{equation}
F_{\alpha\beta} = \sum_i \sum_{k_j}
\frac{N_{\rm modes}^{(i)}(k_j)}{2}\;
\frac{\partial\ln P_0}{\partial\theta_\alpha}\bigg|_{k_j}
\frac{\partial\ln P_0}{\partial\theta_\beta}\bigg|_{k_j},
\label{eq:fisher_desi}
\end{equation}
where $N_{\rm modes} = V_{\rm eff}\,4\pi k^2\delta k/(2\pi)^3$ and
$V_{\rm eff}(k) = V_{\rm survey}[\bar n P_0/(\bar n P_0+1)]^2$.
The sum runs over $k$-bins from $k_{\rm min} = 0.005$ to
$k_{\rm max} = 0.15\,h$~Mpc$^{-1}$, and over all tracer $\times$
redshift bins.  This is a single-spectrum-per-bin Fisher sum: each bin
contributes independently with its own effective volume and bias.
For non-overlapping redshift bins this is exact, but the different
tracer populations overlap in redshift (e.g., LRG and ELG at
$0.6 < z < 1.1$; ELG and QSO at $0.6 < z < 1.6$) and therefore share
the same underlying cosmic modes.  Summing their Fisher contributions as
independent overestimates the available information, since the
cosmic-variance component of the error is correlated.  Because
$\lambda$ is a common multiplicative parameter---it modifies $\Pz(k)$
identically for all tracers---the multi-tracer cosmic-variance
cancellation that benefits parameters like $f_{\rm NL}$ (through
scale-dependent bias differences~\cite{Seljak2009}) does not apply here.
A proper multi-tracer covariance treatment would remove the double
counting without adding compensating information, so our
overlapping-tracer Fisher sum is optimistic and the results should be
regarded as upper bounds on the true DESI sensitivity in the overlapping
redshift range.

We do not include the Alcock--Paczy\'nski effect, higher-order multipoles
($P_2$, $P_4$), or the survey window function.  As discussed in
section~\ref{sec:desi_systematics}, window convolution could suppress the
effective feature amplitude by up to $\sim 50\%$, increasing
$\sigma(\lambda)$ by up to a factor of $\sim 2$.

The parameter set is
$\bm\theta = \{\lambda, a, \kstar, \Dk/\kstar,\,
\ln A_s, n_s,\, b_1,\ldots,b_{N_{\rm bin}}\}$,
where the per-bin biases are nuisance parameters.

We model DESI following the Final Design Report~\cite{DESIFDR2016} and
the 2024 data release~\cite{DESI2024}, with four tracer populations:
BGS ($z < 0.4$), LRG ($0.4 < z < 1.1$), ELG ($0.6 < z < 1.6$), and
QSO ($0.6 < z < 2.1$), covering $f_{\rm sky} \simeq 0.34$ with a
total effective volume $\sim 50\,(\text{Gpc}/h)^3$.  Each tracer class
is subdivided into 3--5 redshift bins (16 bins total).
For comparison, we also compute the Fisher forecast for BOSS
DR12~\cite{BOSS2017} using CMASS ($0.43 < z < 0.70$) and LOWZ
($0.15 < z < 0.43$).

\subsection{Results}
\label{sec:desi_results}

Table~\ref{tab:desi_results} summarizes the sensitivity to $\lambda$
at four levels of marginalization.

\begin{table*}[t]
\centering
\begin{tabular}{lcc}
\hline\hline
Configuration & $\sigma(\lambda)$ & $2\sigma$ threshold \\
\hline
DESI, fixed shape & 0.013 & 0.027 \\
DESI + bias marg. & 0.013 & 0.027 \\
DESI + $\ln A_s$, $n_s$ + bias & 0.013 & 0.027 \\
DESI, full marg.\ (+ SIS shape) & 0.016 & 0.032 \\
\hline
BOSS DR12 (+ $\ln A_s$, $n_s$ + bias) & 0.062 & 0.125 \\
\hline\hline
\end{tabular}
\caption{Preliminary Fisher-matrix sensitivity to the SIS excitation
amplitude $\lambda$ for the fiducial parameters
($\kstar = 0.022\,h$~Mpc$^{-1}$, $\Dk/\kstar = 0.05$, $a = 3$).
These estimates do not include the survey window function (which could
suppress the feature amplitude by up to $\sim 50\%$ and thereby increase
$\sigma(\lambda)$ by up to a factor of $\sim 2$) or marginalization over
shape/distance cosmological parameters beyond $\ln A_s$ and $n_s$.}
\label{tab:desi_results}
\end{table*}

Three features of these results are noteworthy.
First, marginalization over $\ln A_s$ and $n_s$ has essentially no effect
on $\sigma(\lambda)$: the SIS feature is spectrally localized with
oscillatory structure, making it orthogonal to the smooth
$k$-dependence of the primordial amplitude and tilt.  Per-bin bias
marginalization is likewise negligible because bias rescales $P_g(k)$
uniformly.  We caution, however, that our cosmological parameter set is
limited to $\{\ln A_s, n_s\}$; a full treatment including shape and
distance parameters ($\Omega_m h^2$, $\Omega_b h^2$, $h$, neutrino
mass, Alcock--Paczy\'nski dilation) could introduce additional
degradation, particularly through parameters that affect the broadband
$P(k)$ shape near $\kstar$.
Second, full marginalization over the SIS shape parameters
$(a, \kstar, \Dk/\kstar)$ degrades $\sigma(\lambda)$ by only
$\sim 20\%$ (from $0.013$ to $0.016$), because the oscillatory template
is sufficiently distinctive that $\lambda$ is not strongly degenerate
with the shape parameters.
Third, BOSS DR12 gives $\sigma(\lambda) = 0.062$---a factor of
$\sim 5$ weaker than DESI---explaining quantitatively why the SIS
feature was not detected in existing galaxy-survey data.

As a rough indication of the potential improvement from combining CMB
and LSS information, a naive inverse-variance combination of the
indicative Planck sensitivity ($\sigma_{\rm Planck} \approx 0.025$,
from the matched-filter analysis of section~\ref{sec:upper_limits}) with
the DESI Fisher estimate ($\sigma_{\rm DESI} = 0.016$) gives
$\sigma_{\rm joint} \approx 0.014$.  This should not be interpreted as a
rigorous joint constraint, since both inputs are approximate: the Planck
number is an indicative matched-filter estimate, not a posterior from the
full likelihood, and the DESI number does not include window-function
effects or a complete cosmological parameter set.

The DESI sensitivity is approximately independent of the superoscillatory
parameter $a$ (varying by $< 15\%$ for $1.5 \leq a \leq 10$), in
contrast to the CMB where increasing $a$ exponentially suppresses the
oscillatory signal through transfer-function smearing.  This makes galaxy
clustering a uniformly sensitive probe across the full superoscillatory
parameter range.  The sensitivity improves for wider features
($2\sigma \lesssim 0.01$ for $\Dk/\kstar \gtrsim 0.15$) and for
features at higher $k$ ($2\sigma \lesssim 0.01$ for
$\kstar \gtrsim 0.05\,h$~Mpc$^{-1}$), where more $k$-modes are
available.

\begin{figure*}[t]
\centering
\includegraphics[width=\textwidth]{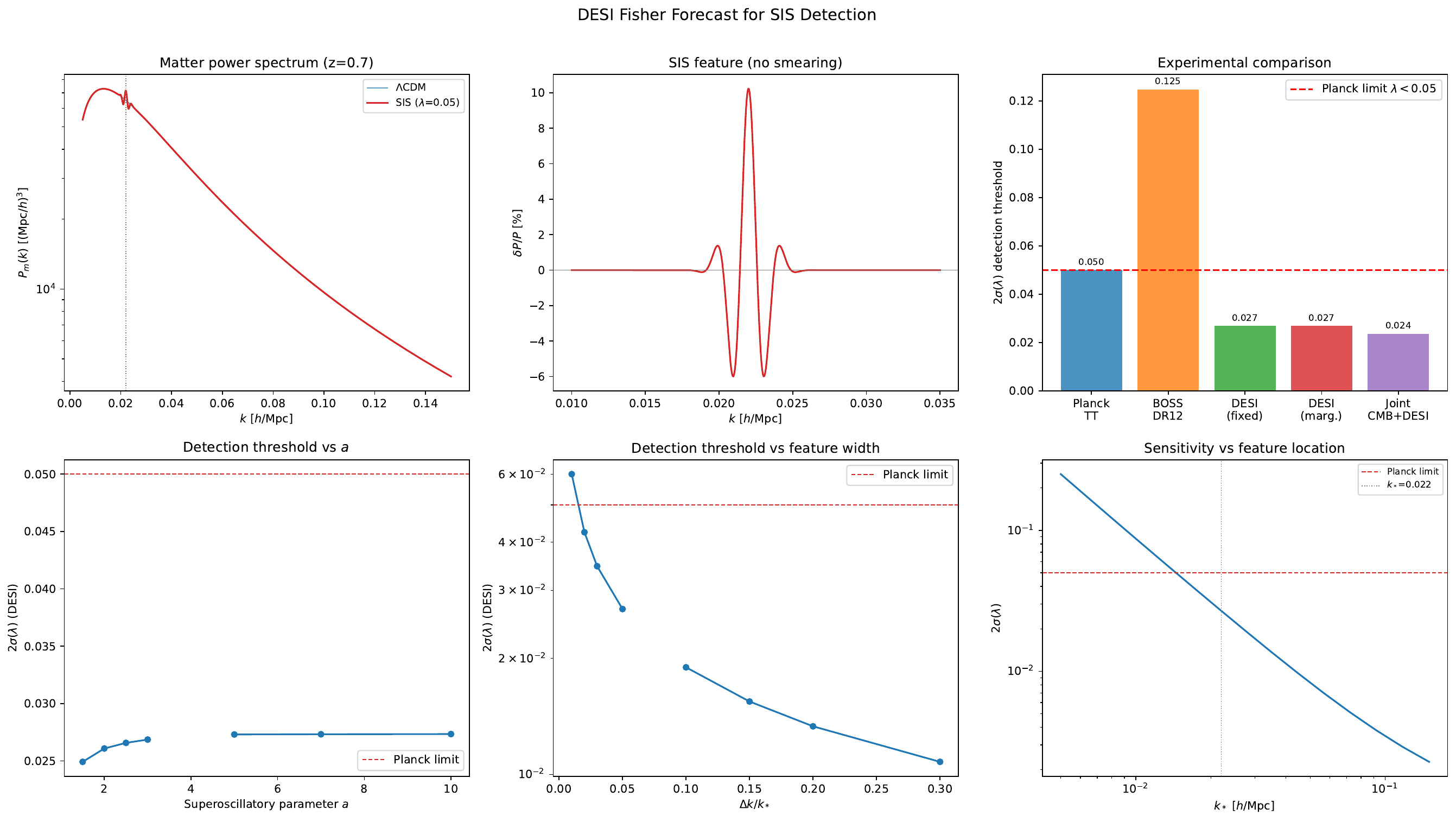}
\caption{DESI Fisher forecast for SIS detection.  Top left: matter power
spectrum at $z = 0.7$ with the SIS feature.  Top center: the SIS
feature in $\delta P/P$, preserved without transfer-function smearing.
Top right: experimental comparison of $2\sigma$ detection thresholds.
Bottom row: DESI sensitivity vs superoscillatory parameter $a$ (left),
feature width $\Dk/\kstar$ (center), and feature location $\kstar$
(right).  The Planck indicative bound $\lambda \lesssim 0.05$ is shown as
the red dashed line.}
\label{fig:desi}
\end{figure*}

\subsection{Systematic considerations}
\label{sec:desi_systematics}

The DESI survey window function has characteristic width
$\delta k_W \sim 0.001$--$0.002\,h$~Mpc$^{-1}$, comparable to the SIS
feature width, and could partially smooth the signal.  We estimate
the amplitude suppression from window convolution to be at most
$\sim 50\%$, based on the ratio $\delta k_W/\Dk$; since the Fisher
information scales as the square of the template amplitude, this would
increase $\sigma(\lambda)$ by up to a factor of $\sim 2$.  A proper
treatment requires convolving the SIS template with the DESI window
function and including the resulting mode-coupling in the covariance
matrix.  Fiber
collisions primarily affect $k \lesssim 0.01\,h$~Mpc$^{-1}$, below the
feature scale.  BAO wiggles ($\Delta k_{\rm BAO} \sim 0.06\,h$~Mpc$^{-1}$)
are $\sim 60\times$ broader than the SIS feature and cleanly separable.
Multiplicative systematics (calibration, extinction) are broadband and
absorbed by the bias and cosmological parameter marginalization.

\section{Discussion and outlook}
\label{sec:discussion}

The principal value of the SIS framework lies in the theoretical
machinery it provides and, as demonstrated in
section~\ref{sec:desi}, in its testability with current data.  The
boundary-action construction of section~\ref{sec:boundary} shows
that a single quadratic boundary term at the initial time surface---a
minimal modification of the inflationary action---is sufficient to
produce a rich and highly structured primordial spectrum.  The Bogoliubov
coefficients, the oscillatory ringing, the smooth quadratic bump, the
Gaussian dominance condition, and the backreaction bound all emerge from
one object: the boundary kernel $\kappa_k$.  This economy of structure is
a direct inheritance from the mathematics of superoscillations, which
shows that band-limited functions can encode arbitrarily rapid local
variations through quantum interference among a finite set of harmonics.

At the same time, the present implementation is intentionally
phenomenological.  The boundary kernel is chosen within the boundary-EFT
language to realize a localized superoscillatory excitation, and the
central scale $\kstar$ is scanned observationally rather than derived
from a microscopic model.  The robust predictions of the framework are
therefore not the absolute location of the feature, but the correlated
structure that follows once $(a,N,\kstar,\Dk,\lambda)$ are specified:
the oscillatory and smooth components of the power spectrum, the TT/EE
ratio, the bispectrum coefficients, and the band-limited consistency
relations.  Deriving the SIS kernel and the preferred scale $\kstar$ from
an explicit UV completion remains an important open direction.

The resulting phenomenology is not a free parametrization but
a tightly constrained framework.  The five structural constraints of
section~\ref{sec:distinction}---the superoscillatory frequency condition,
the width ratio, the minimum-order bound $N > a^2-1$, the binomial
harmonic spectrum, and the exponential bin-averaging suppression---are
all consequences of band-limitedness and together form a highly
restrictive, overdetermined pattern not expected in generic excited-state
ans\"atze.  Any detection of a localized oscillatory feature
in the primordial spectrum would face a specific, overdetermined set of
consistency checks that either confirm or rule out a superoscillatory
origin.

The CMB analysis reveals an observability window
$1 < a \lesssim 3$--$4$ set by the transfer-function smearing, and the
indicative Planck matched-filter bound $\lambda \lesssim 0.05$ already
lies in a regime where further power-spectrum improvements are likely to
be modest: TT is close to cosmic-variance limited and the oscillatory
component is transfer-smeared.

The galaxy clustering forecast of section~\ref{sec:desi} changes this
picture.  The three-dimensional power spectrum preserves the full SIS
oscillatory structure without transfer-function smearing, and
DESI's large survey volume provides sufficient spectral resolution to
marginally resolve the feature.  The preliminary Fisher-matrix analysis
yields $\sigma(\lambda) = 0.016$ after full marginalization over
$\ln A_s$, $n_s$, per-bin galaxy bias, and the SIS feature shape---a
$2\sigma$ threshold of $\lambda \simeq 0.032$ before accounting for
systematic effects.  However, the survey window function could suppress
the effective feature amplitude by up to $\sim 50\%$; since the Fisher
information scales as the square of the template amplitude,
$\sigma(\lambda)$ would increase by up to a factor of $\sim 2$, pushing
the $2\sigma$ threshold to $\sim 0.06$---at or above the indicative
Planck bound.  In addition, the forecast marginalizes only over $\{\ln A_s,
n_s\}$ and per-bin biases, not over shape and distance parameters
($\Omega_m h^2$, $h$, etc.) that a full galaxy-clustering analysis
would require.  Whether DESI can robustly probe below the indicative
Planck bound therefore depends on a more complete forecast incorporating
the survey window function, a broader cosmological parameter set, and
the actual DESI data covariance.

One notable structural feature of the forecast is that marginalization
over $\ln A_s$ and $n_s$ has no effect on $\sigma(\lambda)$: the
spectrally localized, oscillatory SIS template is orthogonal to these
smooth $k$-dependent parameters.  Whether this orthogonality persists
after including shape parameters that modify the broadband $P(k)$ near
$\kstar$ remains to be verified.  The DESI sensitivity is also
approximately independent of the superoscillatory parameter $a$---in
contrast to the CMB, where increasing $a$ exponentially suppresses the
signal---making galaxy clustering a uniformly sensitive probe across the
full superoscillatory parameter range.

The main contributions of this work are:
\begin{enumerate}
\item An explicit boundary-action construction linking superoscillatory
quantum interference to inflationary initial-state modifications---the
first such derivation in the inflation literature.
\item A transfer-function smearing calculation, calibrated against full
Boltzmann computation, establishing the CMB observability window
$1 < a \lesssim 3$--$4$ and revealing that the analytic Gaussian
approximation overestimates the signal by a factor of $\sim 3$.
\item Correlated predictions for TT and EE spectra, including the
identification of the TT/EE modification ratio as a diagnostic of the
feature width.
\item Bispectrum predictions with specific coefficients ($c_1 = 3$ for
equilateral, $c_2 = 1/2$ for folded), obtained from semi-analytic
evaluation of the in-in integral and showing that the dominant
non-Gaussian signal is a spectrally localized folded bispectrum with
$|f_{\rm NL}^{\rm fold}| \sim \lambda/(2\epsilon)$.
\item Five structural constraints distinguishing SIS from generic excited
states, arising from band-limitedness and providing an overdetermined
consistency check for any future detection.
\item A preliminary Fisher-matrix forecast indicating that DESI galaxy
clustering offers a qualitatively different probe of the SIS signal---free
of transfer-function smearing---with $\sigma(\lambda) = 0.016$ before
accounting for the survey window function and with a limited cosmological
parameter set.  Whether DESI can robustly probe below the indicative
Planck bound remains an open question requiring a more complete forecast.
\end{enumerate}

The most immediate extensions are: (i)~a full likelihood analysis of
Planck unbinned TT data at multipole resolution $\delta\ell \sim 1$;
(ii)~a joint TT+TE+EE analysis exploiting the TT--EE width diagnostic;
(iii)~an analysis of DESI data using the SIS matched-filter template,
properly accounting for the survey window function and observational
systematics; (iv)~forecasts for future surveys
(Euclid, SPHEREx, Roman), which will provide even larger effective volumes;
and (v)~a microscopic derivation of the SIS boundary kernel from a
UV-complete inflationary model.

The SIS framework shows that the mathematics of quantum
superoscillations translates naturally into a complete and falsifiable
inflationary phenomenology.  The boundary-action construction provides
a bridge between the foundational quantum mechanics of Aharonov and
Berry and the observational program of precision cosmology.  Whether
nature has chosen a superoscillatory initial state is an empirical
question; what this work provides is a demonstration that such a choice would
leave a distinctive, tightly constrained imprint on the universe, and that
galaxy clustering offers a qualitatively new channel---free of
transfer-function smearing---for searching for such an imprint.
Whether current surveys have the sensitivity to probe below the
indicative Planck bound is a question that a dedicated analysis,
properly incorporating the survey window function, multi-tracer
covariance, and a complete cosmological parameter set, can answer.

\section*{Acknowledgments}

I thank Yakir Aharonov and Michael Berry for inspiring discussions on
quantum superoscillations.

\appendix
\raggedbottom

\section{Wronskian normalization and boundary-condition derivation}
\label{app:wronskian}

We verify the Wronskian normalization $u_k u_k'^* - u_k^* u_k' = i$
used in the derivation of eq.~\eqref{eq:beta_from_kernel}, and provide
intermediate steps for the boundary-condition algebra.

The Bunch--Davies mode function and its derivative are
\begin{align}
u_k(\tau) &= \frac{1}{\sqrt{2k}}\left(1 - \frac{i}{k\tau}\right)
e^{-ik\tau},
\\
u_k'(\tau) &= \frac{e^{-ik\tau}}{\sqrt{2k}}\left(-ik - \frac{1}{\tau}
+ \frac{i}{k\tau^2}\right).
\end{align}
Computing the products:
\begin{align}
u_k\,u_k'^* &= \frac{1}{2k}\left(1-\frac{i}{k\tau}\right)
\left(ik - \frac{1}{\tau} - \frac{i}{k\tau^2}\right)
\nonumber\\
&= \frac{1}{2k}\left[ik - \frac{1}{k^2\tau^3}\right],
\label{eq:uu_star}
\end{align}
where all intermediate terms cancel in pairs.  Similarly:
\begin{equation}
u_k^*\,u_k' = \frac{1}{2k}\left[-ik - \frac{1}{k^2\tau^3}\right].
\label{eq:u_star_u}
\end{equation}
Subtracting:
\begin{equation}
u_k\,u_k'^* - u_k^*\,u_k'
= \frac{1}{2k}\bigl[2ik\bigr] = i.
\label{eq:Wronskian_check}
\end{equation}

\paragraph{Denominator of eq.~\eqref{eq:beta_from_kernel}.}
Using $\kappa_k^{\rm BD} = u_k'/u_k$:
\begin{equation}
u_k'^*(\tau_*) - \kappa_k^{\rm BD}\,u_k^*(\tau_*)
= \frac{u_k\,u_k'^* - u_k^*\,u_k'}{u_k(\tau_*)}
= \frac{i}{u_k(\tau_*)}.
\label{eq:denom_BD_app}
\end{equation}
Including the deformation $\delta\kappa_k$:
\begin{equation}
u_k'^*(\tau_*) - \kappa_k\,u_k^*(\tau_*)
= \frac{i}{u_k(\tau_*)} - \delta\kappa_k\,u_k^*(\tau_*).
\label{eq:denom_full_app}
\end{equation}
Dividing the numerator $-\delta\kappa_k\,u_k(\tau_*)$ by this denominator
and multiplying through by $u_k(\tau_*)$ yields
eq.~\eqref{eq:beta_from_kernel}.

\section{Transfer-function convolution}
\label{app:convolution}

We derive the smearing formula~\eqref{eq:dCl_osc} in detail.  The SIS
correction to $C_\ell^{TT}$ is
\begin{equation}
\delta C_\ell = \frac{2}{\pi}\int \de k\;k^2\,\PBD(k)\,\delta(k)\,
|\Delta_\ell^T(k)|^2,
\label{eq:dCl_integral_app}
\end{equation}
where $\delta(k) = 2\lambda\,G(k)\cos\Phi(k) + \lambda^2 G^2(k)$
with $G(k) = e^{-(k-\kstar)^2/(2\Dk^2)}$ and
$\Phi(k) = a(k-\kstar)/\Dk$.

Modeling the squared transfer function at fixed $\ell$ as a Gaussian
centered at $k_\ell = \ell/D_*$ with width $\sigma_k$:
\begin{equation}
|\Delta_\ell^T(k)|^2 \simeq \mathcal{N}_\ell\,
\exp\!\left[-\frac{(k-k_\ell)^2}{2\sigma_k^2}\right],
\label{eq:transfer_gaussian}
\end{equation}
the product of the two Gaussians (from $G(k)$ and the transfer kernel)
gives a new Gaussian:
\begin{equation}
G(k)\,|\Delta_\ell^T(k)|^2 \propto
C_{12}\;\exp\!\left[-\frac{(k-\bar k)^2}{2\sigma_{12}^2}\right],
\label{eq:product_gaussian}
\end{equation}
with
\begingroup
\setlength{\abovedisplayskip}{4pt}
\setlength{\abovedisplayshortskip}{0pt}
\setlength{\belowdisplayskip}{4pt}
\setlength{\belowdisplayshortskip}{4pt}
\begin{equation}
\begin{aligned}
\sigma_{12}^2 &= \frac{\sigma_k^2\Dk^2}{\sigma_k^2+\Dk^2},
\qquad
\bar k = \frac{k_\ell\Dk^2 + \kstar\sigma_k^2}{\sigma_k^2+\Dk^2}, \\
C_{12} &= \exp\!\left[-\frac{(k_\ell-\kstar)^2}{2(\sigma_k^2+\Dk^2)}\right].
\end{aligned}
\label{eq:convolution_params}
\end{equation}
\endgroup
The remaining integral over $\cos[\Phi(k)]$ evaluates to
\begin{multline}
\int \de k\;e^{-(k-\bar k)^2/(2\sigma_{12}^2)}\;
\cos\!\left[\frac{a(k-\kstar)}{\Dk}\right] \\
= \sqrt{2\pi}\,\sigma_{12}\,
\exp\!\left[-\frac{a^2\sigma_{12}^2}{2\Dk^2}\right]
\cos\!\left[\frac{a(\bar k-\kstar)}{\Dk}\right],
\label{eq:fourier_gaussian}
\end{multline}
using the standard Fourier transform of a Gaussian.  Substituting
$\bar k - \kstar = (k_\ell - \kstar)\Dk^2/(\sigma_k^2+\Dk^2)
= (\ell-\ell_*)\Dk^2/[D_*(\sigma_k^2+\Dk^2)]$ and collecting
factors yields eq.~\eqref{eq:dCl_osc} with the smearing
parameters~\eqref{eq:ell_eff}--\eqref{eq:smearing_factor}.


\end{document}